\documentclass[%
 reprint,
nofootinbib,
 amsmath,amssymb,
 aps,
]{revtex4-2}

\newcommand{\jcap}{J.\ Cosmol.\ Astropart.\ Phys.}

\def\apj{ApJ}
\def\aj{AsJ}
 
\def\apjs{ApJS}               
\def\prd{Phys.~Rev.~D}
\def\physrep{Phys.~Rep.}
\def\aap{A\&A}

\newcommand{\mnras}{Mon.\ Not.\ R.\ Astron.\ Soc.}

\newcommand{\rhokhatI}{\rho^{\hat{\kappa} I}}
\newcommand{\rhokhatg}{\rho^{\hat{\kappa} g}}
\newcommand{\rhogI}{\rho^{gI}}
\newcommand{\Clkhatg}{ C_l^{\hat{\kappa}g}}
\newcommand{\ClkhatI}{ C_l^{\hat{\kappa}I}}
\newcommand{\Clkhatdelg}{ C_l^{\hat{\kappa}^{\mathrm{cln}}g}}

\usepackage{graphicx}
\usepackage[export]{adjustbox}
\usepackage{subfig}
\usepackage{dcolumn}
\usepackage{bm}
\usepackage{xcolor}

\begin{document}

\preprint{APS/123-QED}

\title{Model independent variance cancellation in CMB lensing cross-correlations}
\author{Antón Baleato Lizancos}
\email{a.baleatolizancos@berkeley.edu}

\author{Simone Ferraro}%
\email{sferraro@lbl.gov}

\affiliation{Berkeley Center for Cosmological Physics, Department of Physics, University of California, Berkeley, CA 94720, USA}%
\affiliation{Lawrence Berkeley National Laboratory, One Cyclotron Road, Berkeley, CA 94720, USA}

\date{\today}

\begin{abstract}
Cross-correlations of CMB lensing reconstructions with other tracers of matter constrain primordial non-Gaussianity, neutrino masses and structure growth as a function of cosmic time. We formalize a method to improve the precision of these measurements by using a third tracer to remove structure from the lensing reconstructions. Crucially, our method enjoys the variance reduction benefits of a joint-modelling approach without the need to model the cosmological dependence of the ancillary tracer. We present a first demonstration of variance cancellation using data from Planck and the DESI Legacy Surveys, showing a 10--20\% reduction in both lensing power and cross-correlation variance using the Cosmic Infrared Background (CIB) or DESI Legacy Survey Luminous Red Galaxies (LRGs) as matter tracers.
\end{abstract}

\maketitle


\section{Introduction}
The cosmic microwave background (CMB) is the oldest light we can observe; it is made up of photons which (for the most part) last scattered at redshift $z \approx 1100$. The CMB we see has been gravitationally lensed by the distribution of matter -- both luminous and dark -- that the photons encountered along their trajectory, an effect that can be harnessed to reconstruct maps of that very matter distribution in projection (see~\cite{ref:lewis_challinor_review} for a review).

These reconstructions can in turn be cross-correlated with other tracers of matter to extract insights that cannot be gleaned with either tracer alone. This is one of the most promising ways to measure the growth of cosmic structures, primordial non-Gaussianity, or the sum of the neutrino masses~\cite{ref:schmittfull_and_seljak}. Moreover, cross-correlations make it possible to isolate contributions from different redshifts, a prized property in times of tantalizing discrepancies between probes of early and late cosmic times (see, e.g.,~\cite{Abdalla:2022yfr, ref:white_et_al_22, Krolewski:2021yqy, DES:2021wwk, Heymans:2020gsg} and references therein). Heuristically, if $\hat{\kappa}$ is a reconstruction of the CMB lensing convergence\footnote{We write the reconstruction as $\hat{\kappa}$ to differentiate it from the true $\kappa$. Note, however, that $g$ and $I$ are noisy observations.} and $g$ is some other tracer of the matter distribution -- such as a galaxy survey -- with redshift support $z_{g}$, the `clumpiness' of matter at the time corresponding to $z_{g}$ can be determined from a ratio of angular spectra, $\sigma_8(z_{g})\sim C^{\hat{\kappa}g} / \sqrt{C^{gg}}$, where $\sigma_8$ refers to the amplitude of the linear matter power spectrum on a scale of $8\,\mathrm{h}^{-1}$\,Mpc \cite{ref:yu_et_al_18, Yu:2021vce}. Similarly, galaxy bias, including any scale dependence induced by primordial non-Gaussianity, can be extracted from $b(z_{g})\sim C^{gg}/C^{\hat{\kappa}g}$. For a typical tracer $g$, its auto-correlation is measured much more accurately than its cross-correlation with lensing, so the uncertainty on $\sigma_8(z_{g})$ and $b(z_{g})$ is dominated by the error on $C^{\hat{\kappa}g}$.

Reducing the cross-correlation error requires limiting chance correlations between features in the lensing and galaxy maps. One way to account for these is to introduce a third tracer, $I$, which correlates with structures in $\hat{\kappa}$ that are not correlated with $g$, and modeling everything jointly (e.g.,~\cite{ref:schmittfull_and_seljak}). Often times, however, we might be interested in obtaining constraints that are independent of $I$, be it because the tracer cannot be modelled easily or accurately, or to avoid introducing a dependence on physics from the cosmic era sourcing $I$.

In this work, we explore an alternative approach to variance reduction that relies solely on measurable quantities and limits the dependence on redshifts different from those we intend to isolate. Our method entails subtracting a filtered version of $I$ from $\hat{\kappa}$ in a procedure analogous to `delensing'. In section~\ref{sec:theory}, we derive the optimal form of these filters and forecast potential gains in signal-to-noise. Then, in section~\ref{sec:demo}, we demonstrate that the variance reduction seen on real data from Planck and the DESI Legacy Surveys matches theoretical expectations. In an appendix, we generalize our method to the case where $g$ and $I$ are correlated, explaining how to account for this in real analyses. 

\section{Theory}\label{sec:theory}
Consider $\hat{C}_l^{\hat{\kappa} g}$, the measured angular cross-correlation of $\hat{\kappa}$ and $g$. In general, $\hat{C}_l^{\hat{\kappa} g}$ follows a $\chi^2$ distribution with $2l+1$ degrees of freedom, but away from the lowest few multipoles the distribution can be approximated as Gaussian by the central limit theorem. In this regime, the variance of the $i$th multipole bin of $\hat{C}_l^{\hat{\kappa} g}$ is given by
\begin{equation}\label{eqn:var}
    \sigma^{2}\left( \hat{C}^{\hat{\kappa} g}_i \right) = \frac{1}{(2l_i+1)f_{\mathrm{sky}}\Delta l} \left[C_i^{\hat{\kappa} \hat{\kappa}} C_i^{gg} + \left(C_i^{\hat{\kappa} g}\right)^2\right]\,,
\end{equation}
where $f_{\mathrm{sky}}$ is the fraction of sky covered by the observations, $\Delta l$ is the width of the bins (which we assume to be uniform for simplicity), and $l_i$ is the central multipole of the $i$th bin. This expression suggests that, by removing structure from $\hat{\kappa}_{lm}$, we can suppress $C_i^{\hat{\kappa} \hat{\kappa}}$ and thus lower the variance of the measurement\footnote{It follows from equation~\eqref{eqn:var} that removing the correlated part also leads to lower variance, though the gain is typically subdominant to that stemming from a reduction in $C_i^{\hat{\kappa}\hat{\kappa}}$.}. 

With this goal in mind, let us introduce a third tracer, $I$, which is partially correlated with $\hat{\kappa}$ (and possibly also with $g$). This can be any tracer of low redshift matter such as a the cosmic infrared background (CIB) or a galaxy density or weak lensing (shear) field. We can use it to obtain a `redshift cleaned' convergence map as\footnote{The same effect can be achieved by including $f_l C_l^{\hat{\kappa}I}$ in the likelihood.}
\begin{equation}\label{eqn:map_level_delensing}
    \hat{\kappa}^{\mathrm{cln}}_{lm} = \hat{\kappa}_{lm} - f_{lm}I_{lm}\,,
\end{equation}
where $f$ is a filter to be optimized shortly; throughout this work, we will assume that $I$ is statistically homogeneous such that the optimal $f$ is isotropic (i.e., independent of $m$), but the method applies more generally. 

Recently, refs.~\cite{ref:mccarthy_et_al_21, ref:qu_et_al_22, Maniyar:2021arp} determined the weights that approximately null contributions from a given redshift range. Let us instead determine the choice of $f$ that maximizes the signal-to-noise on a measurement of the cross-correlation. The signal-to-noise ratio of the $i$th bin is defined as
\begin{equation}
    \left(S/N\right)_i \equiv \frac{C_i^{\hat{\kappa} g}}{\sigma\left(\hat{C}_i^{\hat{\kappa} g}\right)} = \sqrt{\frac{(2l_i+1)f_{\mathrm{sky}}\Delta l}{1+\left(\rhokhatg_i\right)^{-2}}}\,,
\end{equation}
where $\rhokhatg$ is the correlation coefficient between the lensing reconstruction, $\hat{\kappa}$, and tracer $g$, and is defined as\footnote{This is not to be confused with the correlation between $g$ and the true lensing convergence, $\rho^{\kappa g}_i = \rhokhatg_i \left(1 + N^{\kappa\kappa}_i/C^{\kappa\kappa}_i\right)^{1/2}$, where $N^{\kappa\kappa}$ is the power spectrum of the reconstruction noise.} 
\begin{equation}
    \rhokhatg_i = \frac{C_i^{\hat{\kappa} g}}{\sqrt{C_i^{\hat{\kappa}\hat{\kappa}}C_i^{g g}}}\,,
\end{equation}
where we take $C_i^{\hat{\kappa}\hat{\kappa}}$ and $C_i^{g g}$ to include reconstruction noise and shot noise, respectively. It follows that the choice of $f_l$ that maximizes the signal-to-noise is also that which maximizes $\rho_l^{\hat{\kappa}^{\mathrm{cln}} g}$.

There are two effects we must consider when maximizing $\rho^{\hat{\kappa}^{\mathrm{cln}} g}$ with respect to $f$: on the one hand, any non-zero field we add to $\hat{\kappa}$ will affect the variance (i.e., `noise') of $\hat{\kappa}^{\mathrm{cln}}$; on the other, if tracers $g$ and $I$ are correlated, the cross-correlation signal will itself be impacted. Taking both into account, we determine the optimal filter to be
\begin{equation}\label{eqn:mv_weights}
    f_l = \frac{C_l^{\kappa I}}{C^{II}_l}\left(\frac{\rho_l^{\hat{\kappa} g} - \rho_l^{g I}/\rho_l^{\hat{\kappa} I}}{\rho_l^{\hat{\kappa} g} - \rho_l^{g I}\rho_l^{\hat{\kappa} I}}\right) \equiv  \frac{C_l^{\kappa I}}{C^{II}_l} \gamma_l\,,
\end{equation}
granted $\rhogI_l\neq1$ (when $\rhogI_l = 1$, $g$ and $I$ are one and the same tracer, and one must logically set $f_l=0$). Note that $C_l^{II}$ includes all sources of noise. If tracers $g$ and $I$ are completely uncorrelated, $\rhogI_l=0$, so $\gamma_l=1$ and $f\rightarrow C^{\kappa I}_l/C^{II}_l$ -- a form familiar from several applications of variance reduction in cosmology~\cite{ref:sherwin_15, ref:qu_et_al_22}.

The form of $f_l$ given in equation~\eqref{eqn:mv_weights} guarantees an improvement in the S/N of the cross-correlation. However, the goal of this work is to achieve this while \emph{removing} structure in $\hat{\kappa}$ that correlates with $I$; this sets the additional requirement that $f_l>0$ (c.f. equation~\ref{eqn:map_level_delensing})\footnote{Assuming $I$ is positively correlated with $\kappa$. The converse holds when they are anticorrelated, as would be the case for a map of the tSZ effect below 217\,GHz, for example.}. It can be shown that $f_l>0$ if and only if
\begin{equation}\label{eqn:VC_criterion}
    \rhokhatI_l > \frac{\rhogI_l}{\rhokhatg_l}
     \quad \text{or}\quad \rhokhatI_l > \frac{\rhokhatg_l}{\rhogI_l}\,.
\end{equation}
When $\rho_l^{gI}=0$, the first condition is met automatically. On the other hand, when $\rho_l^{gI}\neq0$ the situation is more nuanced, as the $\rhokhatI_l$-$\rhokhatg_l$ plane splits into regions where either one or none of the above conditions are satisfied: for reference, a prerequisite for the first condition to be met is that $\rhokhatg_l>\rhogI_l$; for the second, the proviso is $\rhokhatg_l<\rhogI_l$. We study this general case in detail in appendix~\ref{appendix:generalization}.

Though both of these qualitatively-different scenarios in principle allow for variance cancellation that results in improved S/N, they differ in their practical benefits. When $\rho_l^{gI}\neq 0$, our efforts to cancel variance will to some extent entail a reduction of the signal, making modeling difficult unless the removed signal can be accounted for accurately enough; moreover we will have introduced an undesired dependence on tracer $I$. In appendix~\ref{appendix:generalization}, we show that when $\rhokhatI_l > \rhogI_l/\rhokhatg_l$, measurement uncertainties are typically small enough that the removed signal can either be ignored or restored based on direct measurements of $ \hat{C}_l^{gI}$, with the benefits from variance cancellation outweighing the additional error introduced when characterizing this correction term empirically. However, when $\rhokhatI_l > \rhokhatg_l/\rhogI_l$ the penalty from restoring the signal is higher, and values of $\{\rhokhatI_l, \rhokhatg_l, \rhogI_l\}$ that both satisfy $\rhokhatI_l > \rhokhatg_l/\rhogI_l$ and lead to overall improved precision are in most cases only possible if $ \hat{C}_l^{gI}$ can be measured on a patch of sky larger than the one where we carry out the redshift cleaning. In fact, it is true more generally that the balance of factors -- and thus the change in S/N after cleaning -- depends on the relative sizes of these two patches. 

\begin{figure}
    \centering
    \includegraphics[width=\columnwidth]{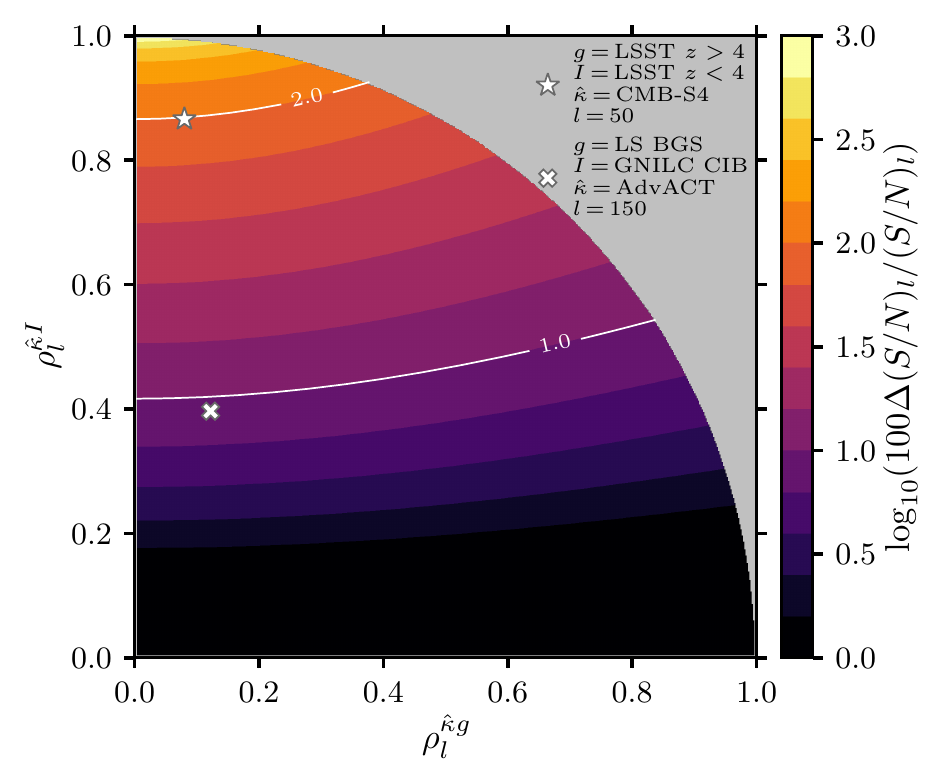}
    \caption{Fractional change in the signal-to-noise ratio per mode of $\hat{C}_l^{\hat{\kappa} g}$ after cleaning $\hat{\kappa}$ with an optimally filtered tracer $I$, in the limit that $\rho_l^{g I}=0$. The grey region corresponds to values of $\rho_l^{\hat{\kappa} g}$ and $\rho_l^{\hat{\kappa} I}$ incompatible with $\rho_l^{g I}=0$ (see appendix~\ref{appendix:bound_on_rho}). The star and cross represent two particular combinations of data outlined in the legend and described in the main text.} \label{fig:change_in_StoN}
\end{figure}

Leaving the caveat of signal restoration to appendix~\ref{appendix:generalization}, the fractional change in signal-to-noise per mode of $\Clkhatg$ that can be attained using our optimal weights is 
\begin{widetext}
\begin{align}\label{eqn:delta_S_to_N}
    \frac{\Delta (S/N)_l} {(S/N)_l} =\left\{1 + \left[1+\left(\rho_l^{\hat{\kappa} g}\right)^2\right]^{-1}\left[\frac{1+\gamma_l(\gamma_l-2)\left(\rho_l^{\hat{\kappa} I}\right)^2}{\left(1-\gamma_l \rho_l^{g I}\rho_l^{\hat{\kappa} I}/\rho_l^{\hat{\kappa} g} \right)^2}-1\right]\right\}^{-\frac{1}{2}} - 1\,.
\end{align}
\end{widetext}
Figure~\ref{fig:change_in_StoN} shows the improvement in signal-to-noise as a function of $\rho_l^{\hat{\kappa} g}$ and $\rho_l^{\hat{\kappa} I}$ in the limit that $g$ and $I$ are uncorrelated. Improvements of several tens of percent are achievable with realistic tracers, especially whenever $\rho_l^{\hat{\kappa} I}$ is significantly higher than $\rho_l^{\hat{\kappa} g}$. For a given value of $\rho_l^{\hat{\kappa} I}$, the fractional improvement in S/N is larger the smaller $\rho_l^{\hat{\kappa} g}$ is. This is because with lower $\rhokhatg_l$, the term proportional to $C_l^{\hat{\kappa}\hat{\kappa}}$ dominates the error in equation~\eqref{eqn:var} more clearly, amplifying the impact of cleaning. The method is thus especially suited to improve low-significance measurements or to make first detections. 

To illustrate the promise of redshift cleaning, we consider as a potential application the use of low-redshift ($z<4$) measurements from the future Vera Rubin Observatory (VRO) LSST~\cite{ref:lsst} to clean CMB-S4~\cite{ref:s4_science_book} lensing reconstructions, and subsequently correlating these with the highest-redshift bin of LSST ($z>4$)\footnote{We assume the LSST bins to be disjoint. Further details about our parametrization of the tracers are given in appendix~\ref{appendix:parametrizing_tracers_forecasts}.}. We find that the S/N of the cross-correlation is improved by $\sim100\%$ at $l\sim50$, as shown by the star in figure~\ref{fig:change_in_StoN}.\footnote{Improvements are even larger at lower $l$, but we do not focus on those scales here as they require a more refined treatment due to the Limber approximation breaking down and systematic effects becoming more important.} This gain could be important, for instance, when searching for $f_{\mathrm{NL}}$, the signal of which peaks at high redshifts and large angular scales~\cite{ref:schmittfull_and_seljak}. Since CMB-S4 is overwhelmingly signal dominated at these low $l$s, the only way to improve measurement significance is by getting around the cosmic variance of $\kappa$. Our method does precisely this while bypassing the need to model the low redshift tracers\footnote{Further work is needed to assess the impact of decorrelation between $\kappa$ and $I$ due to scale-dependent bias and non-linearities.}.

Another promising application is to improve measurements of $\sigma_8$. With existing data, the gains are modest: the cross in figure~\ref{fig:change_in_StoN} shows a $10\%$ gain in precision of $C^{\hat{\kappa}g}_{l=150}$ when using GNILC CIB to clean AdvACT DR6 and cross-correlate with Legacy Survey BGS (assuming $\rho^{gI}=0$; more details about these tracers to come). However, the figure also shows that improvements of a factor of two or greater on the large-scale amplitude of fluctuations are possible with upcoming experiments. 

\section{Demonstration}\label{sec:demo}
We now present a first demonstration of variance cancellation by cross-correlating CMB lensing data from Planck with galaxy survey data from the DESI Legacy Survey (LS)~\cite{ref:dey_et_al_22}. Since the benefits of our method are predicated on being able to remove true lensing modes that do not correlate with $g$, and the Planck lensing noise levels are relatively high -- only a handful of modes are signal-dominated~\cite{ref:carron_et_al_22_lensing_PR4} -- it is at present difficult to find a combination of data sets for which the S/N improves. However, this will soon change dramatically as data from AdvACT~\cite{ref:advact_16}, SO~\cite{ref:SO_science_paper}, SPT-3G~\cite{ref:spt3g_14}, and CMB-S4~\cite{ref:s4_science_book} become available. It is therefore useful to demonstrate that variance cancellation can be achieved and understood irrespective of whether the S/N improves for the particular scenario under consideration. Hence, and in order to maximize the variance reduction, we set $\gamma_l=1$ in the weights for the remainder of this section.

\subsection{Data}\label{sec:data}
We work with the minimum variance combination of CMB lensing reconstructions obtained from temperature and polarization data from the fourth data release of Planck~\cite{ref:carron_et_al_22_lensing_PR4}. The resulting $\hat{\kappa}$ map covers $\sim67\%$ of the sky and is signal-dominated on scales $10\lesssim L\lesssim70$.

As per large-scale structure tracers, we consider various samples of galaxies photometrically selected from Legacy Survey data for spectroscopic follow-up with DESI. These were spectroscopically-callibrated during DESI survey validation. We define a `BGS' sample as the `bright' ($r<19.5$) subset of targets to be observed by the DESI Bright Galaxy Survey (BGS)~\cite{ref:hahn_et_al_22, ref:myers_et_al_22}. This sample is relatively-low redshift: figure~18 of ref.~\cite{ref:hahn_et_al_22} shows its redshift distribution, which is mostly confined to $z<0.5$. We work also with an `LRG' (Luminous Red Galaxy) sample, using directly the maps provided by ref.~\cite{ref:white_et_al_22}, which are split into four redshift bins with redshift distribution given in their figure~2; this sample is explained in detail in~\cite{ref:white_et_al_22, ref:zhou_et_al_22}. Since the first bin has significant overlap with the BGS sample, we will exclude it whenever we use the LRGs as tracer $I$ and BGS as tracer $g$ in order to limit the correlation between the two samples; the remaining LRG sample spans $0.5< z < 1$. The galaxy maps thus produced cover $\sim 50\%$ and $\sim44\%$ of the sky for BGS and the full LRG sample, respectively, and are largely contained within the Planck $\hat{\kappa}$ footprint. Note that while for now we are restricted to using photometric samples with small but non-zero redshift overlap, spectroscopic data from DESI~\cite{ref:desi_16} and Euclid~\cite{ref:euclid_12} will soon allow us to use samples that are non-overlapping, better ensuring $\rho^{gI} \approx 0$.\footnote{There will be a small contribution from magnification bias; if significant, it could be handled by following appendix~\ref{appendix:generalization}.}

Finally, we work also with the cosmic infrared background (CIB) which is a tracer of star formation and as such originates from a wide range of cosmic times peaking at $z\sim2$. Since this closely matches the CMB lensing kernel, the CIB is highly correlated with CMB lensing (e.g.,~\cite{ref:planck_13_cib, ref:sherwin_15}), making it a paradigmatic candidate to be used as tracer $I$ (as suggested already in~\cite{ref:qu_et_al_22}). Specifically, we use the map extracted from Planck 353\,GHz data using the GNILC algorithm~\cite{ref:gnilc}, masking modes with $l<100$ to avoid contamination from spurious artifacts and residual galactic dust on large angular scales. This overlaps with the $\hat{\kappa}$ patch across $\sim 51\%$ of the sky.

\subsection{Methods}
In order to determine the weights $f_l$, and also to later model the variance reduction, we need fiducial spectra 
for all the auto- and cross- spectra between $\hat{\kappa}$, $I$ and $g$. We follow the approach in~\cite{ref:yu_17} to fitting the spectra involving galaxies, CIB and lensing on the patch where they overlap, but use the \texttt{pyccl} code~\cite{ref:pyccl} and the galaxy redshift distributions given in the LS target selection papers cited above. Note that this does not introduce any unwanted dependence on cosmology or physics at a different redshift, since we do not utilize the best-fit parameters for anything else. The fitting form we use to get $f_l$ need not be the actual physically correct model: all that is required to avoid bias is a smooth fit, and any deviation of the fiducial spectrum from the truth will only result in suboptimal variance reduction~\cite{ref:sherwin_15, ref:yu_17}. Moreover, since the form of $f_l$ we are using was obtained from an optimization procedure, the cleaning performance has no linear-order response to inaccuracies in determining $f_l$.

When coadding LRG maps from different redshift bins to then use them as tracer $I$, we first re-weight the galaxy bins using equation~20 of~\cite{ref:sherwin_15} to better match the CMB lensing kernel and thus maximize the cross-correlation with lensing. Admittedly, this only improves performance marginally for the tracers we consider.

We measure the angular pseudo- cross-spectrum on scales $10<l<1000$ on the same sky patch before and after redfshift cleaning, and estimate the variance on estimates of the bandpowers in each of the two cases from the scatter of the measured $\hat{C}_l$'s within bins of width $\Delta l=90$. When doing this, it is important that the underlying spectrum be flat over the size of the bin. We ensure this by first dividing the $\hat{C}_l$'s by a smooth fiducial spectrum -- the appropriate model for each measurement given in appendix~\ref{appendix:modeling_var} --  and multiply the result by the binned version of this fiducial.

We verify that the pipeline is unbiased by applying it to 1000 Gaussian simulations of all the tracers sharing a degree of correlation that matches what is seen in the data\footnote{See appendix~F of~\cite{ref:cib_delensing_biases} for details on how to generate such simulations.}. The scatter of these outputs gives us our error bars.

\subsection{Results}\label{sec:results}
\begin{figure}
    \centering
    \includegraphics[width=\columnwidth]{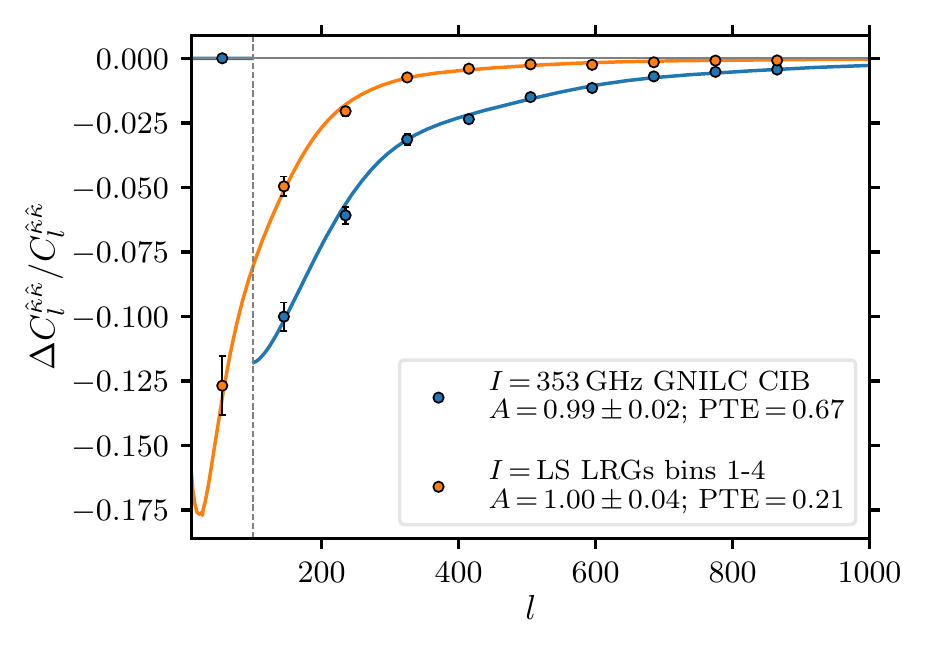}
    \caption{Fractional change in lensing power after cleaning a Planck PR4 $\hat{\kappa}$ map with different filtered tracers (setting $\gamma_l=1$). To reduce sample variance, we bin the numerator only after taking differences between spectra. The data are in excellent agreement with the model (shown as solid lines for a fiducial $A=1$).}
    \label{fig:delta_ckk}
\end{figure}

Figure~\ref{fig:delta_ckk} shows that we are able to effectively remove structure from the lensing maps, and that we do so by an amount that agrees with theoretical expectations. To state this more quantitatively, we fit for an amplitude parameter $A$, which linearly rescales the fiducial model of equation~\eqref{eqn:frac_auto_cls_general} (the solid lines in the figure). Proceeding by least-squares minimization\footnote{Since the bins are wide ($\Delta l =90$), we ignore correlations between them.}, we obtain best-fit parameter values and associated uncertainties  that can be interpreted as highly-significant detections of lensing removal: $41\sigma$ when using GNILC CIB maps to clean Planck PR4 $\hat{\kappa}$, and $25\sigma$ when using LRG bins 1-4. We test goodness-of-fit by calculating the reduced-$\chi^2$ of a binned model with the best-fit value of $A$ and translating this to a probability-to-exceed (PTE); the latter values, quoted in the figure legend, suggest the fits are excellent. Moreover, constraints on $A$ are consistent with the fiducial value of unity.

The reduction is greatest at low $l$ because it is on those large scales that the Planck lensing reconstructions have highest S/N per mode. In addition, whenever we clean with a galaxy map, such as our LRG sample, this is compounded with the fact that it is at low-$l$ that contributions from low redshift structures make up a higher fraction of the total lensing power (see, e.g.,~\cite{ref:lewis_challinor_review}). It is reassuring that below $l<100$, where we masked modes in the CIB map to avoid the impact of systematics, the variance is unchanged.

\begin{figure}
    \centering
    \includegraphics[width=\columnwidth]{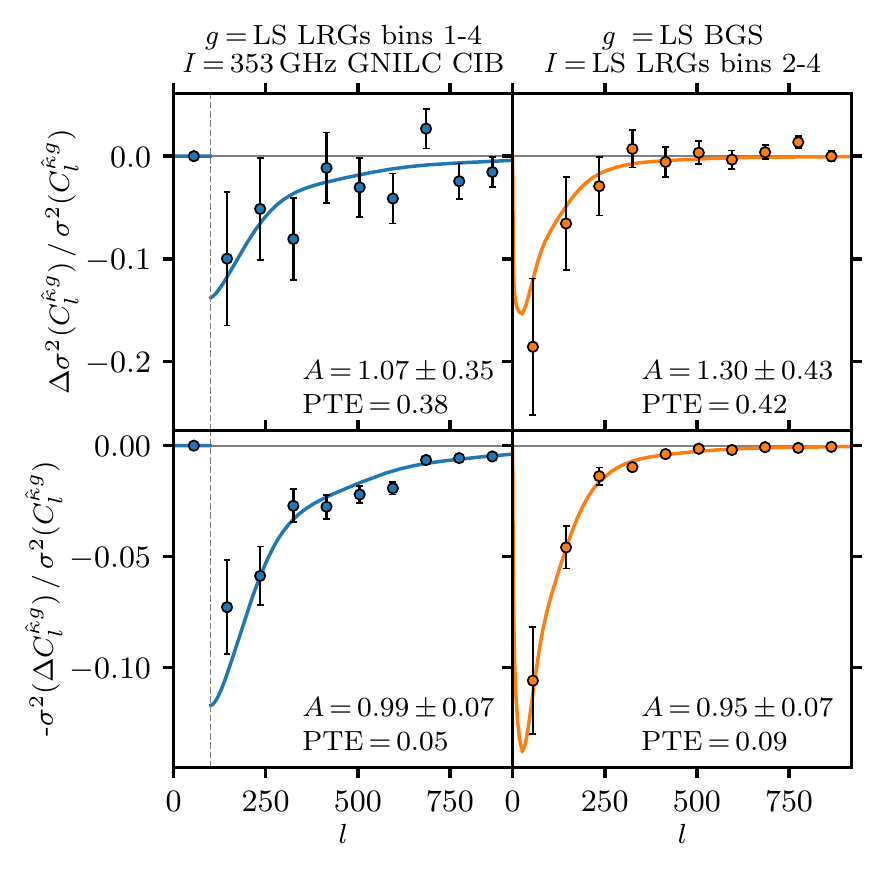}
    \caption{Variance cancellation in the cross-correlation of tracer $g$ with a Planck PR4 $\hat{\kappa}$ map after cleaning the latter with a filtered tracer $I$ (setting $\gamma_l=1$). Top row:  fractional change in the variance per mode. Bottom row: (minus) the variance associated with the structures that cleaning removes, as a fraction of the original variance. According to appendix~\ref{appendix:modeling_var}, top and bottom rows are equal in the mean in the limit that $\rho_l^{gI}=0$, but the former is more noisy. Solid lines show fiducial models with $A=1$, while the best-fit values are annotated in each panel.}
    \label{fig:var_plots}
\end{figure}

This removal of structure from the CMB lensing map translates to lower variance when cross-correlating it with other tracers. In figure~\ref{fig:var_plots}, we demonstrate this for two tracer combinations: in the left column, we clean the Planck $\hat{\kappa}$ map with the GNILC CIB map, and subsequently correlate this with the combination of all four LRG bins; on the right, we clean it with LRG bins 2-4 (recall that this sample spans $0.5<z<1$), weighted to match the lensing kernel, and cross-correlate with the BGS sample ($z<0.5$). 

We consider two estimators of the variance change. The first (top row) shows the fractional difference in variance before and after redshift cleaning. Applying the same significance test as above, we report reductions in the variance of $C_l^{\hat{\kappa}g}$ with confidence slightly above $3\sigma$. This effect is well captured by our fiducial models, which are consistent with the constrained $A$ within $1\sigma$, and yield very plausible PTEs at the best fit values of $A$.

This first estimator makes intuitive sense, but it is rather noisy. In appendix~\ref{appendix:modeling_var}, we derive an alternative one that agrees with the first in the mean when $\gamma_l=1$ and $\rhogI_l=0$ but is less noisy in general; heuristically, it quantifies the variance associated with structures that get removed during the cleaning process. These conditions are not grossly violated in the cases we consider here, particularly in the right column of figure~\ref{fig:var_plots}: $\gamma_l=1$ by design, and though $\rhogI_l$ is strictly speaking not zero, it is small. In this situation, the two estimators differ slightly -- in appendix~\ref{appendix:modeling_var}, we model exactly by how much -- but the second is still a reasonable estimator of the variance change given the statistical errors. Results from this estimator are reported in the bottom two panels, showing a detection significance of approximately $14\sigma$ in the two cases we consider.

For completeness, we show in appendix~\ref{appendix:modeling_var}, figure~\ref{fig:clkk_var_change}, that we can also successfully reduce the variance of the CMB lensing power spectrum by the expected amount, using either the CIB or the LRG samples. We detect this effect with up to $4\sigma$ confidence.


\section{Conclusions}
We have explored a method to reduce cosmic variance in CMB lensing maps by invoking an ancillary tracer which, importantly, does not need to be modelled in detail. This is particularly useful when the tracers are poorly understood, as is the case with the CIB or low-redshift tracers in the non-linear regime. Since our formalism is built around correlation coefficients determined from measurable quantities, it automatically accounts for the gravitational lensing of $g$ and $I$, as well as their decorrelation with each other and with $\kappa$ due to gravitational non-linearities and non-linear bias. With some minor modifications, it is likely that the method could be applicable to galaxy weak lensing as well.

We identified simple conditions that need to be met for the technique to result in improvements in the S/N of CMB lensing cross-correlations. These are automatically satisfied when $\rhogI_l=0$; hence, the method promises to be useful when spectroscopic data are available. On the other hand, when $\rhogI_l\neq0$, some amount of signal is removed during the process, and any benefits depend on the hierarchy between $\rhogI_l$, $\rhokhatI_l$ and $\rhokhatg_l$ (though when $\rhogI_l\ll1$, the bias from signal suppression is typically negligible). In appendix~\ref{appendix:generalization}, we explored ways to partially restore the removed signal, should it be needed, including direct measurements of $\hat{C}_l^{gI}$. When the correlation between $g$ and $I$ is significant, the method is especially suited to enhance deep CMB lensing measurements limited to compact sky areas, especially when $\hat{C}_l^{gI}$ can be measured over larger regions. Users interested in the technique can simply evaluate equations~\eqref{eqn:delta_S_to_N_minimizing_MSE_joint} or~\eqref{eqn:delta_S_to_N_minimizing_MSE_disjoint} to assess whether benefits are accessible to them.

We then demonstrate that variance cancellation can already be achieved using existing tracers. First of all, we report a reduction in Planck PR4 CMB lensing power with $41\sigma$ confidence when using GNILC CIB, or $25\sigma$ using LS LRGs. These lead to $4\sigma$ and $2\sigma$ detections of variance reduction in the CMB lensing auto-spectrum. We then demonstrate variance cancellation in the cross-correlation of CMB lensing with galaxy surveys at $3\sigma$ confidence with standard estimators, or $14\sigma$ with a taylor-made estimator. The theoretical framework we develop proves excellent when it comes to faithfully modelling empirical results.

\appendix

\section*{Acknowledgements}
We thank M. White, G. Farren, F. McCarthy, T. Namikawa, F. Qu and E. Schaan for their insightful comments on an early draft. In addition, we are grateful for fruitful conversations with A. Raichoor, R. Zhou, N. Sailer, J. DeRose, N. Weaverdyck and B. Sherwin. ABL is a BCCP fellow at UC Berkeley and Lawrence Berkeley National Laboratory. SF is funded by the Physics Division of Lawrence Berkeley National Laboratory. This research used resources of the National Energy Research Scientific Computing Center (NERSC), a U.S. Department of Energy Office of Science User Facility operated under Contract No. DE-AC02-05CH11231. We have also made use of NASA's Astrophysics Data System, the arXiv preprint server, the Python programming language and packages \textsc{NumPy, Matplotlib, SciPy, AstroPy}, and \textsc{HealPy}~\cite{ref:healpix_paper, ref:healpy_paper}.

This work was carried out on the territory of xučyun (Huichin), the ancestral and unceded land of the Chochenyo speaking Ohlone people, the successors of the sovereign Verona Band of Alameda County.

\section{Generalization}\label{appendix:generalization}
In equation~\eqref{eqn:VC_criterion}, we identified the two conditions which, if met, guarantee that $f_l>0$, meaning that variance cancellation is possible and will in theory result in improved S/N on $C_l^{\hat{\kappa}g}$. When $\rho_l^{gI}=0$, the first condition is always satisfied, but the situation is more nuanced when $\rho_l^{gI}\neq0$.

\begin{figure*}
    \centering
    \subfloat[]{\includegraphics[width=0.3\textwidth]{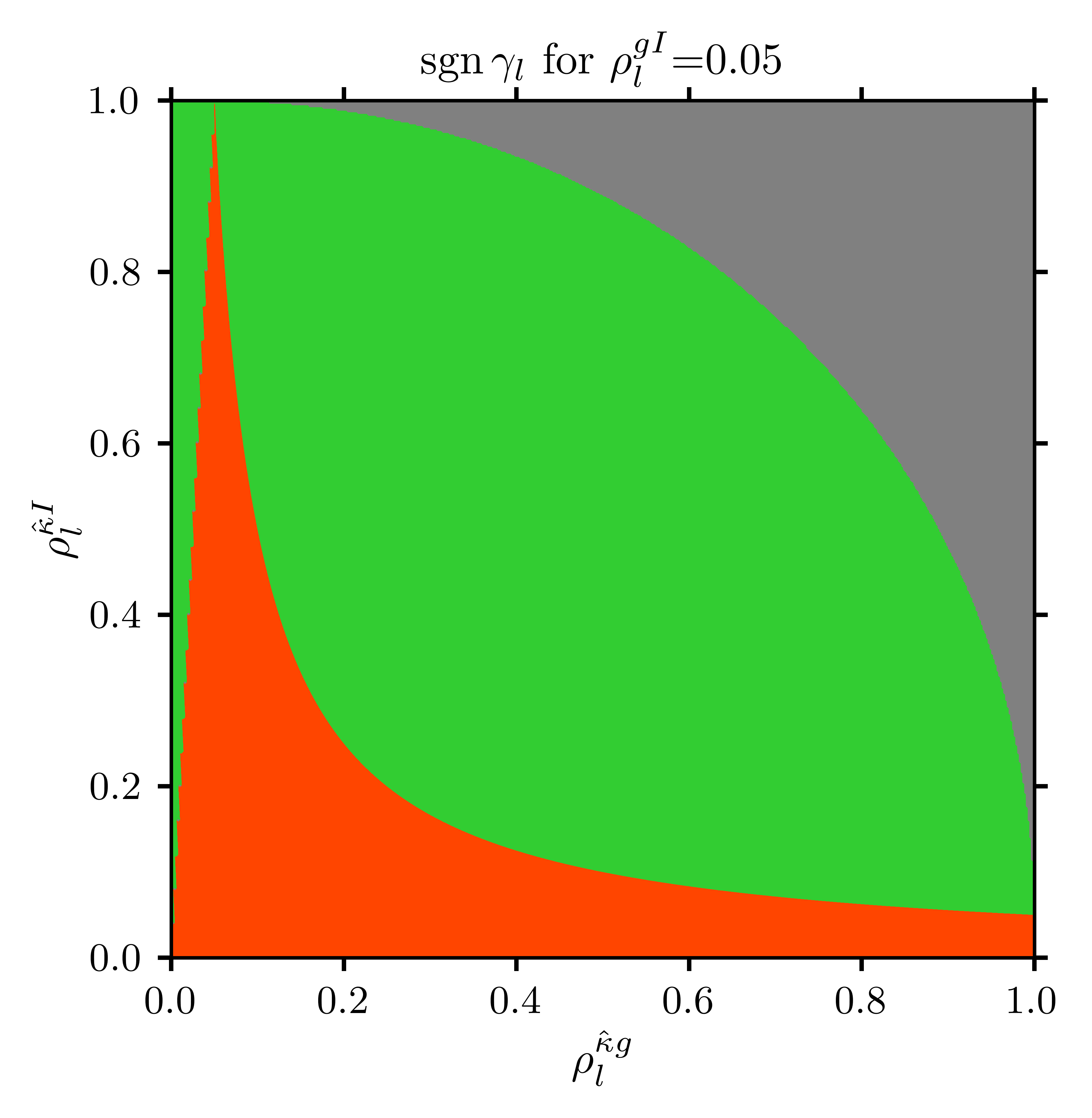}}
    \hfil
    \subfloat[]{\includegraphics[width=0.3\linewidth]{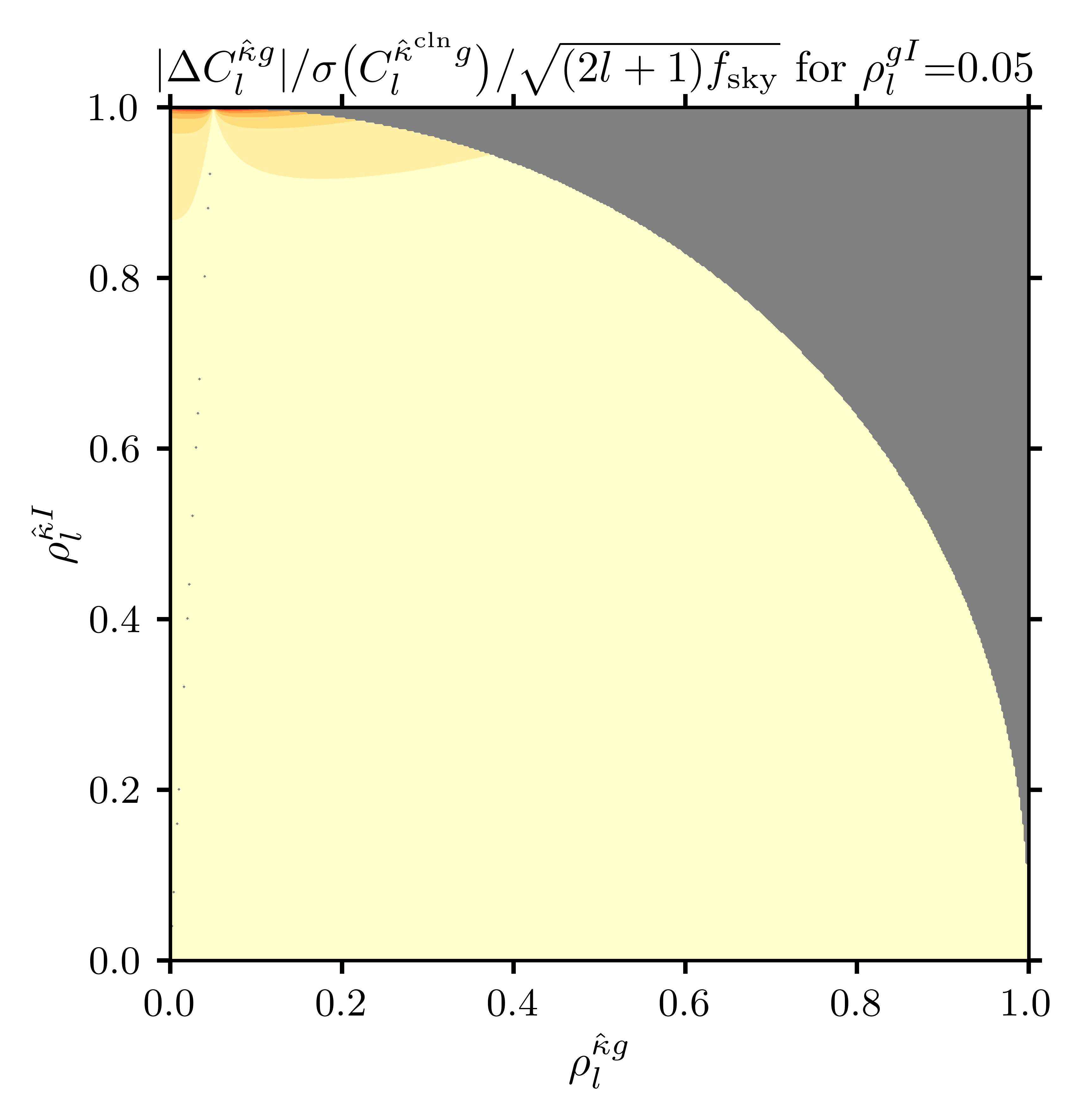}}
    \hfil
    \subfloat[]{\includegraphics[width=0.3\linewidth]{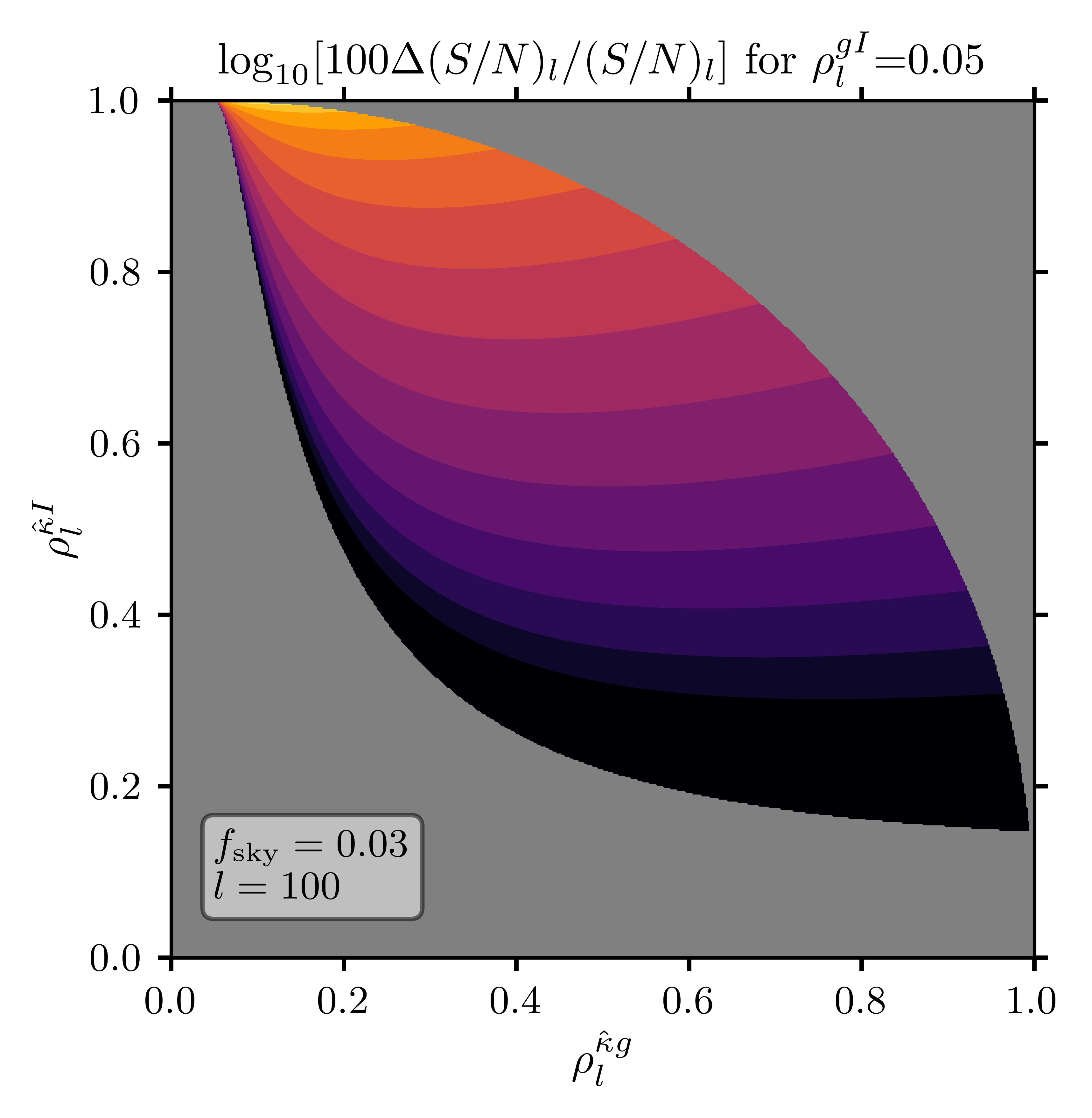}}
    
    \subfloat[]{\includegraphics[width=0.3\textwidth]{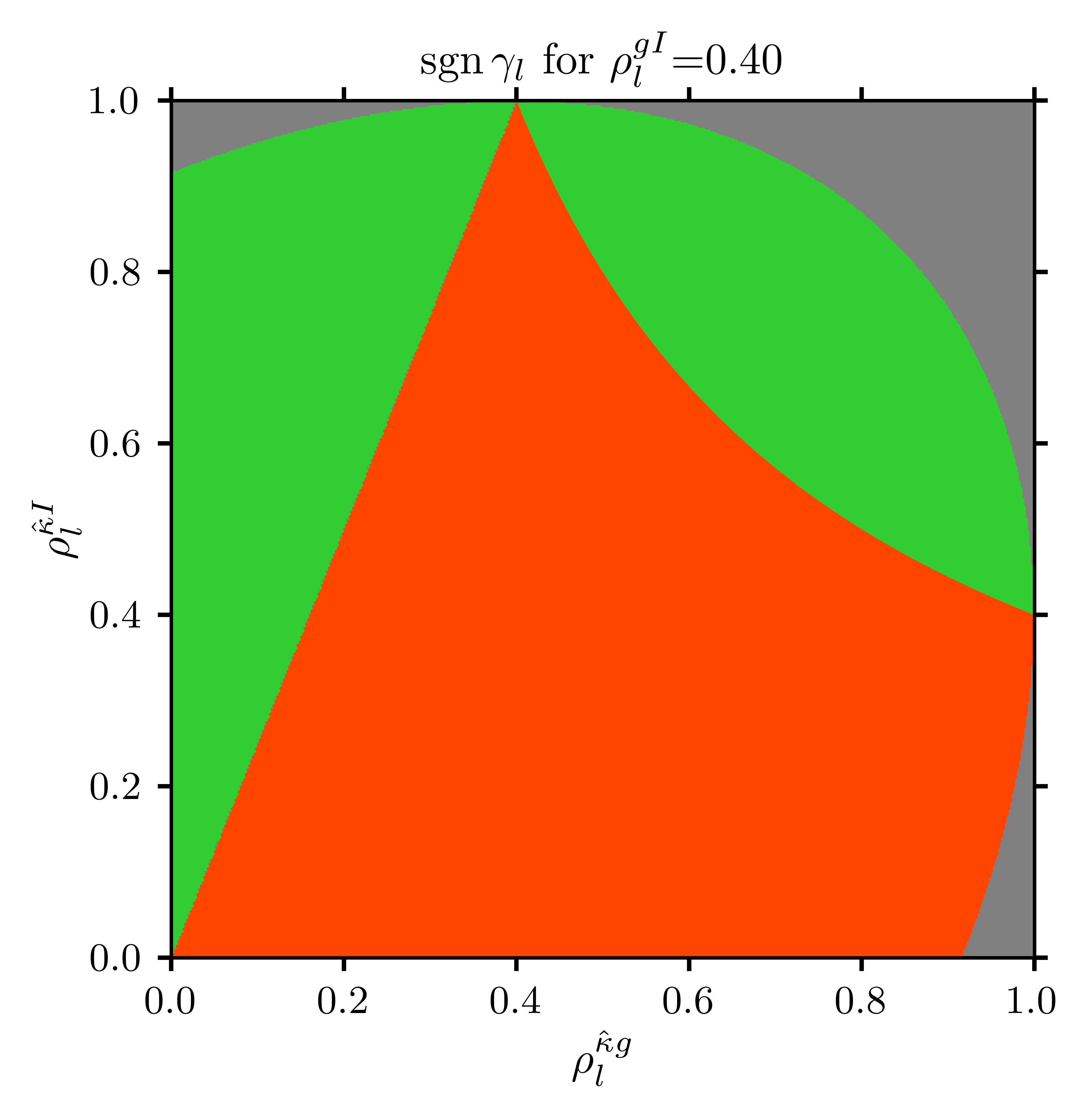}}
    \hfil
    \subfloat[]{\includegraphics[width=0.3\linewidth]{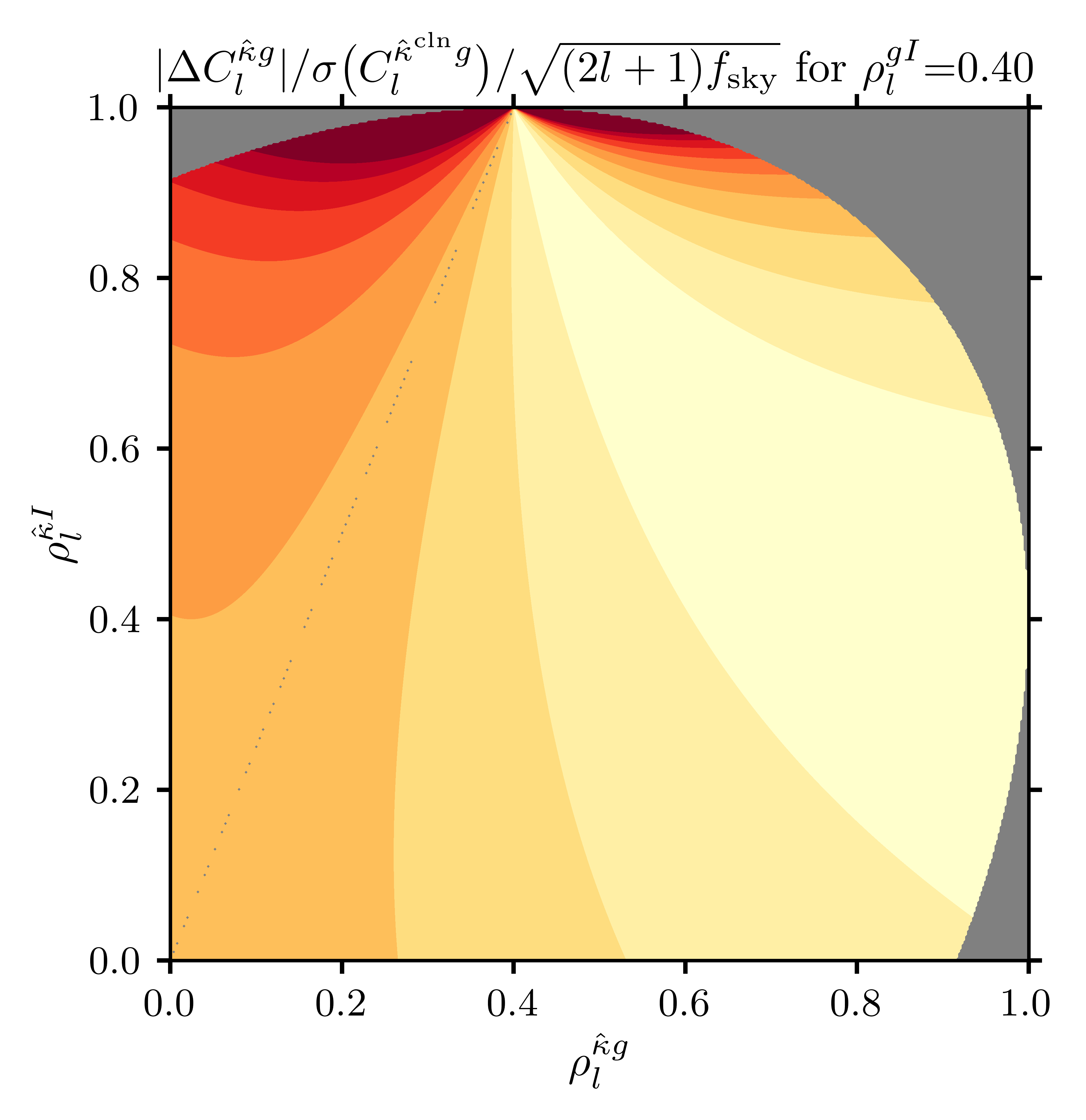}}
    \hfil
    \subfloat[]{\includegraphics[width=0.3\linewidth]{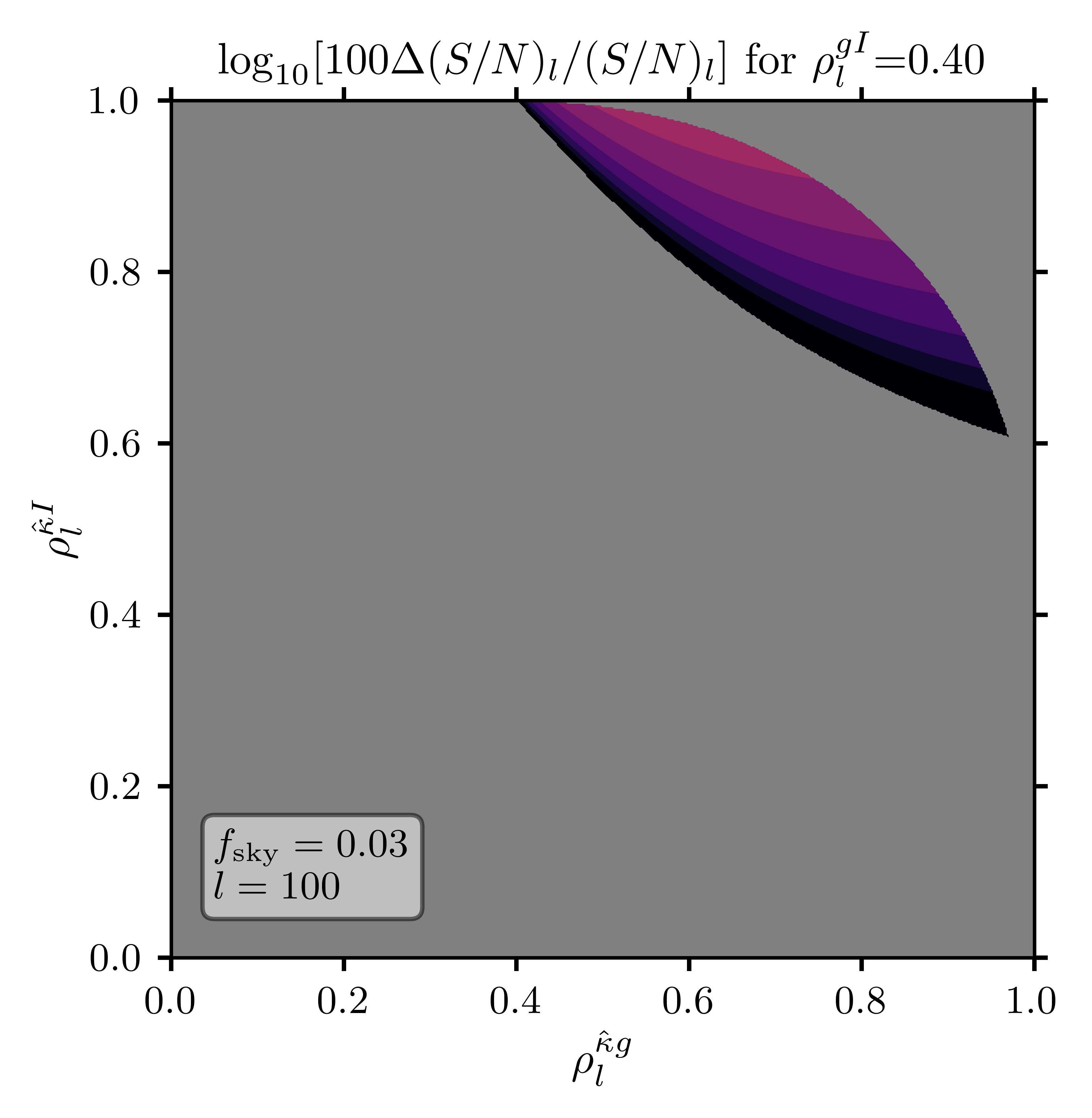}\label{fig:StoN_gain_r=1}}
    
    \subfloat[]{\includegraphics[width=0.3\textwidth]{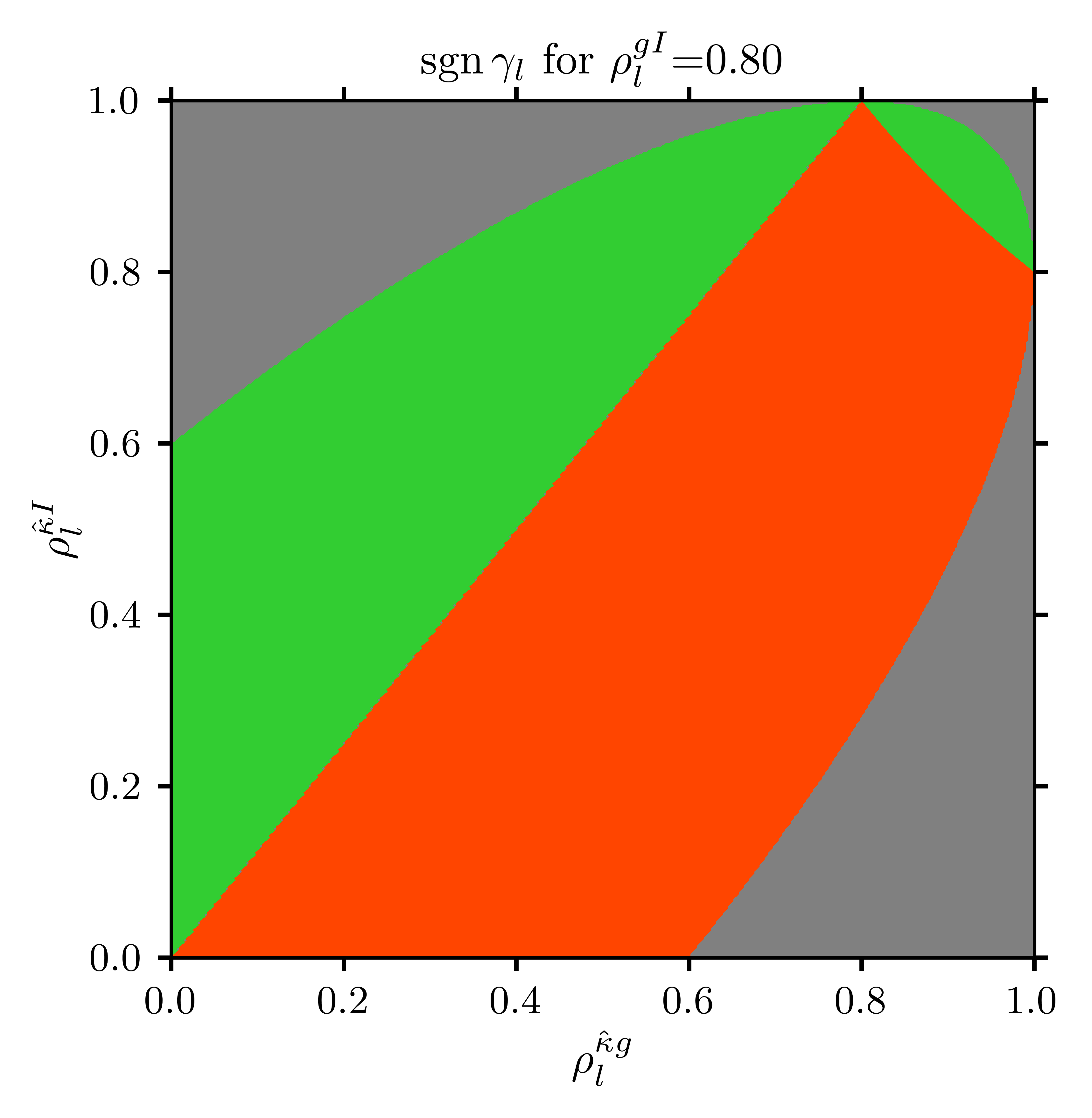}}
    \hfil
    \subfloat[]{\includegraphics[width=0.3\linewidth]{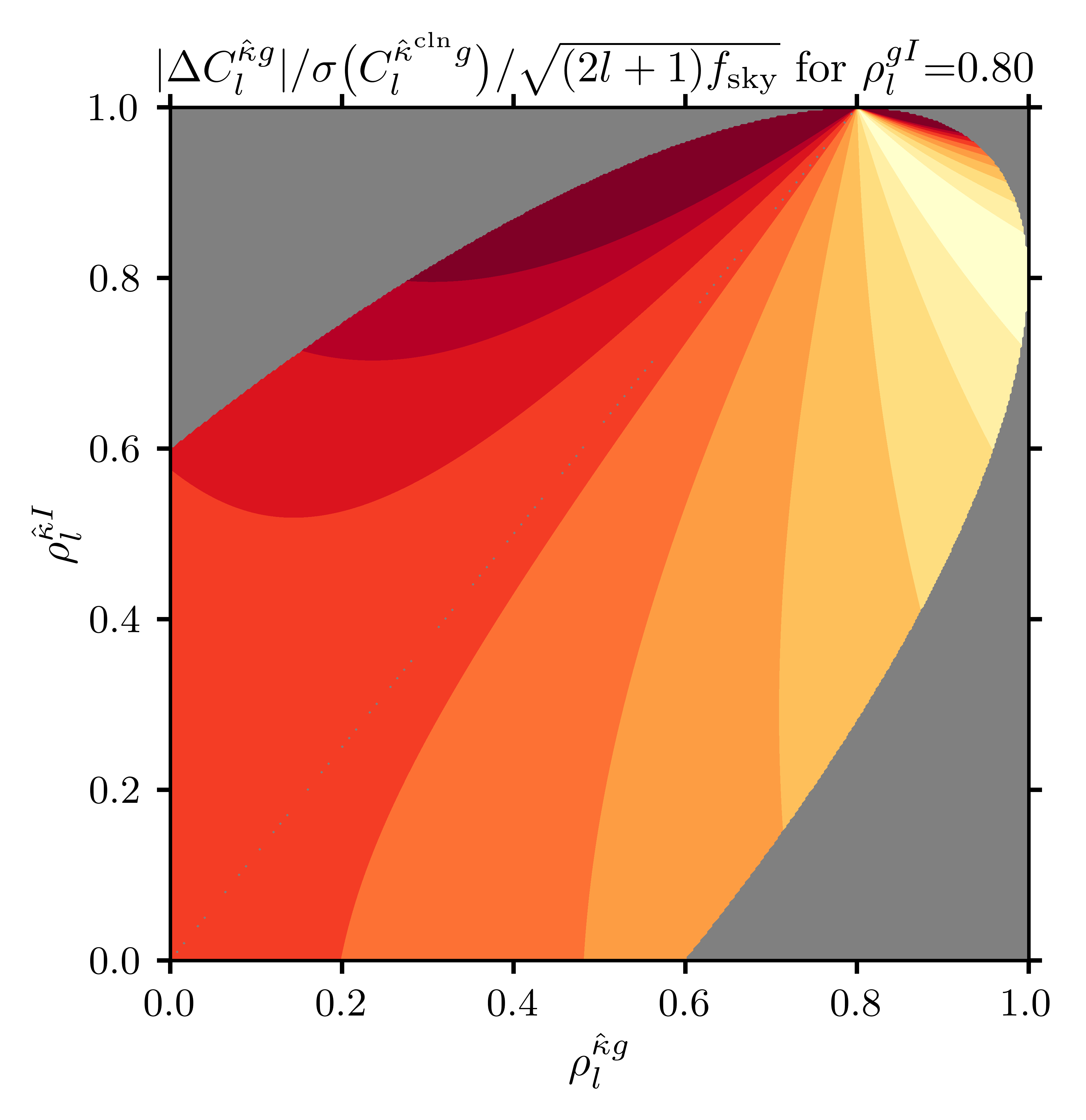}}
    \hfil
    \subfloat[]{\includegraphics[width=0.3\linewidth]{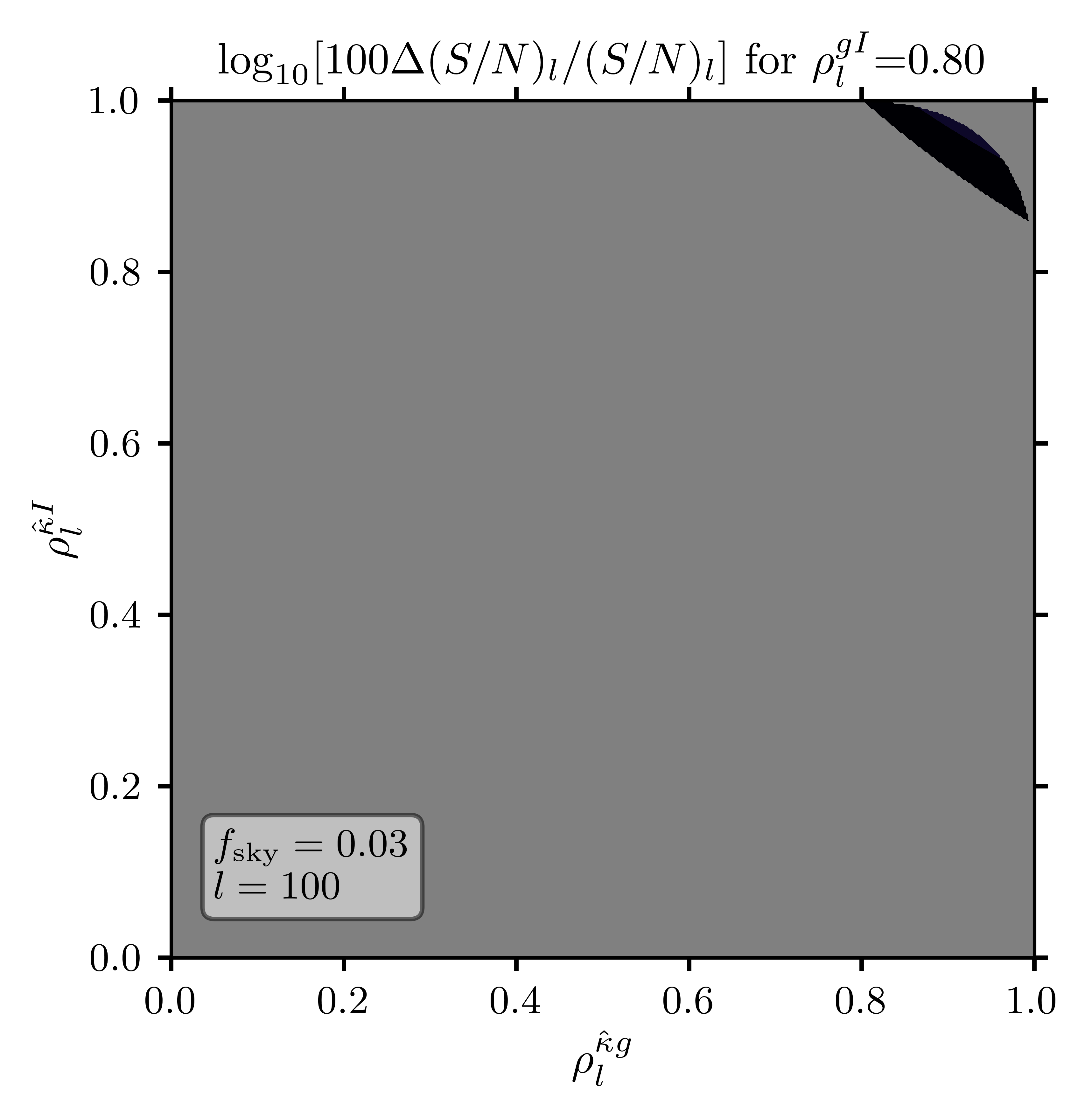}}
    \hfil
    \subfloat[]{\includegraphics[width=0.3\linewidth]{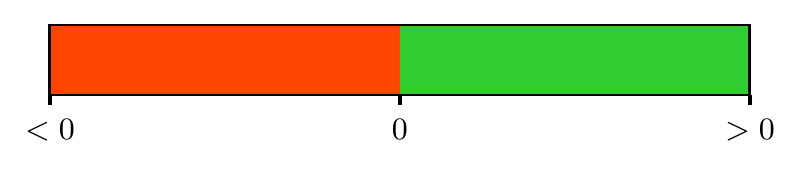}}
    \subfloat[]{\includegraphics[width=0.3\linewidth]{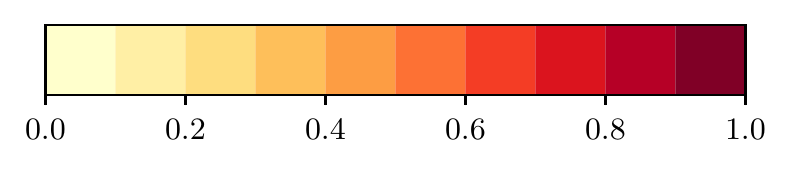}}
    \subfloat[]{\includegraphics[width=0.3\linewidth]{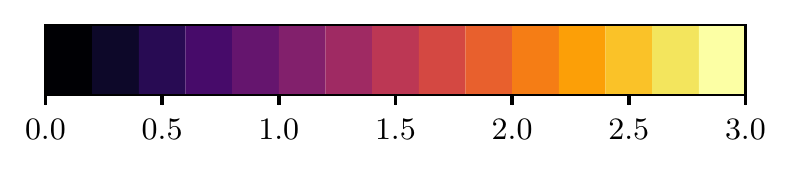}\label{fig:cbar_StoN}}
    \caption{Generalization of the theory of variance cancellation to the case of $\rhogI_l\neq0$. Left column: sign of $f_l$; where positive (green) variance cancellation can in theory lead to improved S/N. However, the two green branches differ by the extent to which the signal is reduced. Middle column: ratio of bias from signal suppression to statistical uncertainty after redshift cleaning. The bias-variance tradeoff can be optimized by minimizing the MSE after a direct measurement of $\hat{C}^{gI}_l$ is used to partially restore the signal. Right panel: fractional change in S/N per mode of $C_l^{\hat{\kappa}g}$ at the MSE optimum after cleaning and restoring the signal from a measurement of $\hat{C}^{gI}_l$ on the same patch, showing only regions where there is a gain in S/N. Animations at intermediate values of $\rhogI_l$ can be found online\footnote{\url{https://abaleato.github.io/kappa_delensing/}}.  }
\label{fig:generalization_w_syst_error}
\end{figure*}

In general, when $\rho_l^{gI}\neq0$, the $\rhokhatI_l$-$\rhokhatg_l$ plane can be divided into regions where $f_l$ is positive or negative. This is illustrated in the left column of figure~\ref{fig:generalization_w_syst_error}: in the green regions, $f_l>0$, and conversely for the red ones. The curves separating these regions are given by the limit where the inequalities in equation~\eqref{eqn:mv_weights} become equalities\footnote{In the discussion that follows, it will be useful to note that the physically allowed region -- as determined in appendix~\ref{appendix:bound_on_rho} --
is tangent to the top axis where $\rhokhatg_l = \rhogI_l$, and to the right axis where $\rhokhatI_l = \rhogI_l$; these are also the points where the thresholds in equation~\eqref{eqn:VC_criterion} intersect the axes.}.

In the regions where $f_l<0$, there can be no gain in S/N from variance cancellation, so equation~\eqref{eqn:map_level_delensing} is essentially telling us to fold $I$ in as a tracer of $g$. In fact, a derivation of $f_l$ using a Lagrange multiplier that constrains $f_l$ to be non-negative returns $f_l=0$ in this regime.

The regions we are interested in, however, are those where $f_l>0$, so we now characterize them in detail. When $\rhokhatg_l>\rhogI_l$, either the first of the two conditions is met -- namely $\rhokhatI_l>\rho_l^{gI}/\rhokhatg_l$ -- or none are. This is telling us that when $g_l$ is more correlated with $\hat{\kappa}_l$ than with $I_l$, it is necessary (though not sufficient) that $I_l$ itself be  more correlated with $\hat{\kappa}_l$ than with $g_l$. Qualitatively, the type of variance cancellation we can obtain in this branch is a natural extension of that we see in the $\rho_l^{gI}=0$ limit. At the dividing line between regions, $\rhokhatI_l=\rho_l^{gI}/\rhokhatg_l$, the weights are zero because this equality can only be satisfied when $\hat{\kappa}_l=I_l$, in which case we are better off not doing anything.

On the other hand, when $\rhokhatg_l<\rhogI_l$, only the second condition -- that is, $\rhokhatI_l>\rhokhatg_l/\rho_l^{gI}$ -- can be satisfied. This time, the necessary (but, again, insufficient) condition is that when $g_l$ is less correlated with $\hat{\kappa}_l$ than with $I_l$, $\hat{\kappa}_l$ must be significantly more correlated with $I_l$ than with $g_l$ (linearly more so).  In this case, the weights are undefined at the boundary between regimes because the one requirement in our derivation of $f_l$ was that $\rhokhatI_l\neq\rhokhatg_l/\rho_l^{gI}$.

These conditions are in fact quite idealized. Both they and equation~\eqref{eqn:delta_S_to_N} tell us about the S/N in a way we cannot directly work with, because when $\rhogI_l\neq0$ we need to know what has happened to the cross-correlation signal through the redshift cleaning process. Whenever tracers $g$ and $I$ are correlated and $f_l>0$, the signal decreases as
\begin{equation}
    C^{\hat{\kappa}^{\mathrm{cln}}g}_l = C^{\hat{\kappa} g}_l - f_{l} C_{l}^{g I}\,.
\end{equation}
As long as the optimal weights in equation~\eqref{eqn:mv_weights} are used, it is in principle still advantageous to pursue variance cancellation. However, attempts to fit the data with a theoretical model for $C^{\kappa g}_l$ will be biased, all the while that the variance is reduced. Moreover, the data will have acquired a dependence on $I$, and thus on physics at redshifts different from that which we would like to isolate.

Let us formalize our definition of this bias as
\begin{align}
    \Delta C^{\hat{\kappa} g}_l & \equiv \langle \hat{C}^{\hat{\kappa}^{\mathrm{cln}}g}_l \rangle - C^{\hat{\kappa} g}_l \nonumber \\
    & = - f_{l} C_{l}^{g I}\,.
\end{align}
The ratio of the bias magnitude to the statistical uncertainty after cleaning is then
\begin{widetext}
\begin{align}\label{eqn:bias_over_noise_uncorrected}
    \frac{|\Delta C^{\hat{\kappa} g}_l|}{\sigma\left(\hat{C}^{\hat{\kappa}^{\mathrm{cln}}g}_l\right)} = \sqrt{(2l+1)f_{\mathrm{sky}}} \left\{ \left(\gamma_l \rho^{gI}_l \rhokhatI_l\right)^{-2}\left[1 + \left(\rho_l^{\hat{\kappa} I}\right)^2 \gamma_l(\gamma_l-2)\right] + \left(\frac{\rhokhatg_l}{\gamma_l \rhogI_l \rhokhatI_l}\right)^{2}\left(1 - \gamma_l \rhogI_l \rhokhatI_l / \rhokhatg_l\right)^{2} \right\}^{-\frac{1}{2}} \,,
\end{align}
\end{widetext}
which is larger the more modes are observed -- hence the factor of $\sqrt{(2l+1)f_{\mathrm{sky}}}$, where $f_{\mathrm{sky}}$ tracks the size of the patch where cleaning is performed (small patches offer greater tolerance to bias due to their higher sample variance). This expression is evaluated in the central column of figure~\ref{fig:generalization_w_syst_error} for three values of $\rhogI_l$, immediately revealing a significant qualitative difference between the two variance cancellation branches identified in equation~\eqref{eqn:VC_criterion}: the fractional bias is smaller when $\rhokhatI_l>\rhogI_l/\rhokhatg_l$ than in the alternative branch where $\rhokhatI_l>\rhokhatg_l/\rhogI_l$. In fact, when $\rhogI_l\ll 1$, the bias can be safely ignored across the former branch for most reasonable values of $l$ and $f_{\mathrm{sky}}$ (as expected, since this is the natural continuation of the $\rhogI_l=0$ case studied in the main text). On the other hand, when $\rhokhatI_l>\rhokhatg_l/\rhogI_l$ the bias is generally larger than the statistical uncertainty except at the largest angular scales and smallest sky patches, for which $\sqrt{(2l+1)f_{\mathrm{sky}}}\lesssim 1$.

If the bias amplitude is unacceptably large for the application at hand, there are ways to mitigate it. The approach we consider here is to actively try to restore the signal we have removed, fitting our theory model to a corrected cross-spectrum
\begin{align}
    \tilde{C}^{\hat{\kappa}^{\mathrm{cln}}g}_l & \equiv \hat{C}^{\hat{\kappa}^{\mathrm{cln}}g}_l + f_{l} C_{l}^{\mathrm{corr}} \,.
\end{align}
For example, if we somehow knew the true $C_l^{gI}$ with no uncertainty, we could set $C_{l}^{\mathrm{corr}} = C_l^{gI}$, and this would eliminate all bias while also retaining all variance suppression. The gain in S/N would then simply be given by equation~\eqref{eqn:delta_S_to_N}. Though $C_l^{gI}$ is unlikely to be known perfectly, it could conceivably be predicted from theory rather accurately if the redshift distributions of the samples are known. Alternatively, it could come from a fit to a wide range of scales and maybe even a larger sky patch, ideally in a way that remains cosmology-independent despite having to assume a fitting form. 

If none of these approaches are viable, $C_l^{gI}$ can still be determined from the data on a multipole-by-multipole basis, but at the cost of increased variance. If we set $C_{l}^{\mathrm{corr}}$ to the measured  $g$-$I$ correlation on the \emph{same} patch where we do the cleaning, $C_{l}^{\mathrm{corr}} = \hat{C}_l^{gI}$, we will have achieved exact unbiasedness but also restored all of the variance we had originally set out to remove.

We therefore seek solutions that lie between the two limits we have seen, compromises between bias and variance reduction\footnote{The metrics we will be introducing in  equations~\eqref{eqn:bias_over_noise} and~\eqref{eqn:MSE} are not adequate in the regime where the optimal weights are negative -- $f<0$, where, heuristically, $I$ is included as a tracer of $g$ to boost the cross-correlation signal. Since the focus of this work is the regime where $f>0$, we show only results appropriate for this case.}. These entail setting $C_{l}^{\mathrm{corr}} = \mathcal{W}_l\hat{C}_l^{gI}$, where $\mathcal{W}_l$ is a filter that we must optimize such that this correction moves us in the direction of unbiasedness but only insofar as we are willing to see the variance increase. One could, for example, choose the form of $\mathcal{W}_l$ that results in the largest amounts of variance reduction subject to the ratio of bias-to-variance, 
\begin{align}\label{eqn:bias_over_noise}
    \frac{|\langle \tilde{C}^{\hat{\kappa}^{\mathrm{cln}}g}_l \rangle - C^{\hat{\kappa} g}_l|}{\sigma\left(\tilde{C}^{\hat{\kappa}^{\mathrm{cln}}g}_l\right)} = \frac{|f_l C_l^{gI}\left(\mathcal{W}_l-1\right)|}{\sigma\left(\hat{C}^{\hat{\kappa}^{\mathrm{cln}}g}_l + f_l\mathcal{W}_l\hat{C}^{gI}_l\right)}\,,
\end{align}
being smaller than some threshold. Note that this expression differs from~\eqref{eqn:bias_over_noise_uncorrected} in that both the numerator and denominator are calculated \emph{after} applying the correction. In particular, the denominator depends on a number of factors, including the size of the patches where cleaning is performed and where $\hat{C}^{gI}_l$ is measured. These need not be the same; if the latter is larger, the uncertainty introduced when restoring the signal will be comparatively smaller; and if the patches are disjoint, the covariance between them will be greatly reduced. In regions where~\eqref{eqn:bias_over_noise_uncorrected} is already below the threshold, the constraint should be inactive and $\mathcal{W}_l=0$: no signal needs to be restored, so variance reduction should be maximal. On the other hand, when the constraint is active, the desired threshold value should be exactly enforced.

Yet another possibility is to choose the form of $\mathcal{W}_l$ that minimizes the mean-squared-error\footnote{Though we do not include it here, the impact of systematic effects in the measurement of $\hat{C}_l^{gI}$, such as inhomogeneities in galaxy samples (e.g.~\cite{ref:huterer_et_al_13, ref:baleato_lizancos_and_white_23}), could be incorporated at this point if its contribution to the error budget after redshift cleaning is known.},
\begin{align}\label{eqn:MSE}
    \mathrm{MSE}\equiv \langle \left( \tilde{C}^{\hat{\kappa}^{\mathrm{cln}}g}_l  - C^{\hat{\kappa} g}_l\right)^2\rangle \,,
\end{align}
thus reducing bias and/or variance according to their relevance in the scenario at hand. This will limit the gains in S/N promised by equation~\eqref{eqn:delta_S_to_N}, but it will also offer the key benefit of limiting the bias amplitude to being at worst of the order of the statistical error. It can be shown that the weights that accomplish this are
\begin{align}\label{eqn:W_weights}
    \mathcal{W}_l = \frac{\left(C_l^{gI}\right)^2 - \mathrm{Cov}\left[\hat{C}^{\hat{\kappa}^{\mathrm{cln}}g}_l|_{\hat{\kappa} \cap g \cap I}, \hat{C}_l^{gI}|_{g \cap I}\right]/f_l }{\left(C_l^{gI}\right)^2 + \sigma^{2}\left(\hat{C}_l^{gI}|_{g \cap I}\right)}\,,
\end{align}
where we have differentiated between measurements made on the patch where $ \hat{C}_l^{gI}$ is measured (which we denote as $ g \cap I$) or in the patch where we carry out the redshift cleaning ($\hat{\kappa} \cap g \cap I$).

When the two patches are one and the same, the weights reduce to
\begin{align}
    \mathcal{W}_l = 1 - &\left[\frac{\rhokhatg_l}{\gamma_l\rhogI_l\rhokhatI_l} + \frac{1}{\gamma_l (\rhogI_l)^2}\right] \nonumber \\
    & \times \left[ 1 + \frac{1}{(\rhogI_l)^2} + (2l+1)f_{\mathrm{sky}}\right]^{-1}\,.
\end{align}
As expected, $\mathcal{W}_l\rightarrow0$ when $\rhogI_l\rightarrow0$, indicating that the bias is best left untreated, as it is small to begin with. At the other extreme, $\mathcal{W}_l\rightarrow1$ as $\rhogI_l\rightarrow1$ -- the bias is so large that mitigating it costs us practically all variance reduction. 

To see this more explicitly, we calculate the change in S/N after minimizing the MSE when the $\hat{\kappa} \cap g \cap I$ and $ g \cap I$ patches are actually the same:
%
\begin{widetext}
\begin{align}\label{eqn:delta_S_to_N_minimizing_MSE_joint}
    \frac{\Delta (S/N)_l} {(S/N)_l} =\left\{1 + \left[1+\left(\rho_l^{\hat{\kappa} g}\right)^2\right]^{-1}\left[\frac{1+\gamma'_l(\gamma'_l-2)\left(\rho_l^{\hat{\kappa} I}\right)^2}{\left(1-\gamma'_l \rho_l^{g I}\rho_l^{\hat{\kappa} I}/\rho_l^{\hat{\kappa} g} \right)^2}-1\right]\right\}^{-\frac{1}{2}} - 1\,,
\end{align}
\end{widetext}
where $\gamma'_l \equiv \gamma_l\left(1-\mathcal{W}_l\right)$; notice that this expression reduces to equation~\eqref{eqn:delta_S_to_N} when $\mathcal{W}_l=0$, and gives zero when $\mathcal{W}_l=1$. The right column of figure~\ref{fig:generalization_w_syst_error} evaluates it for $l=100$ and $f_{\mathrm{sky}}=0.03$, a sky fraction similar to that surveyed by SPT-3G. These values help us illustrate some important features without loss of generality. Note, in particular, that gains in S/N are limited to a subregion of the branch where $\rhokhatI_l>\rhogI_l/\rhokhatg_l$.

\begin{figure}[ht!!!!]
    \centering
    \includegraphics[width=0.32\textwidth]{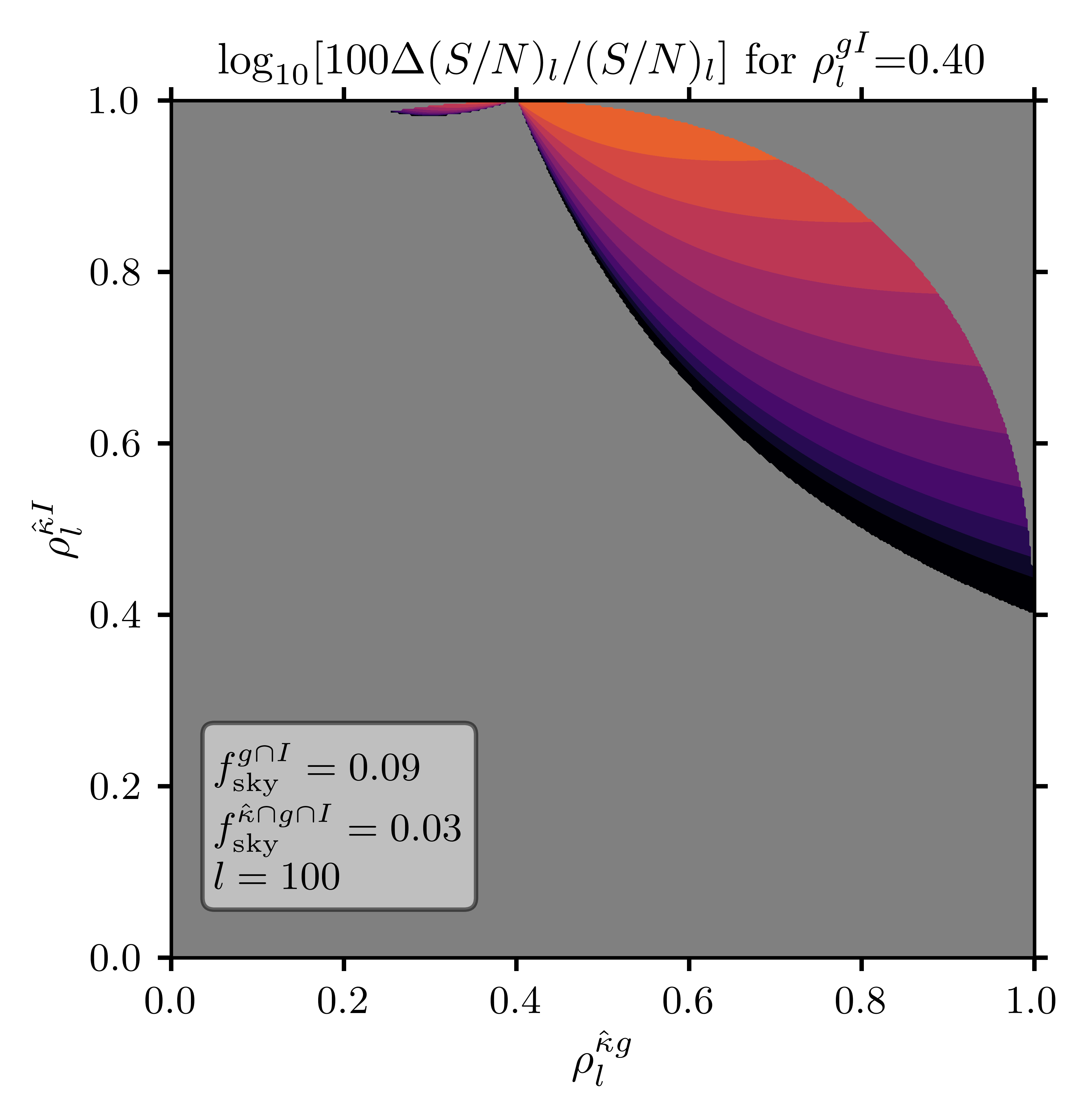}
    \includegraphics[width=0.32\textwidth]{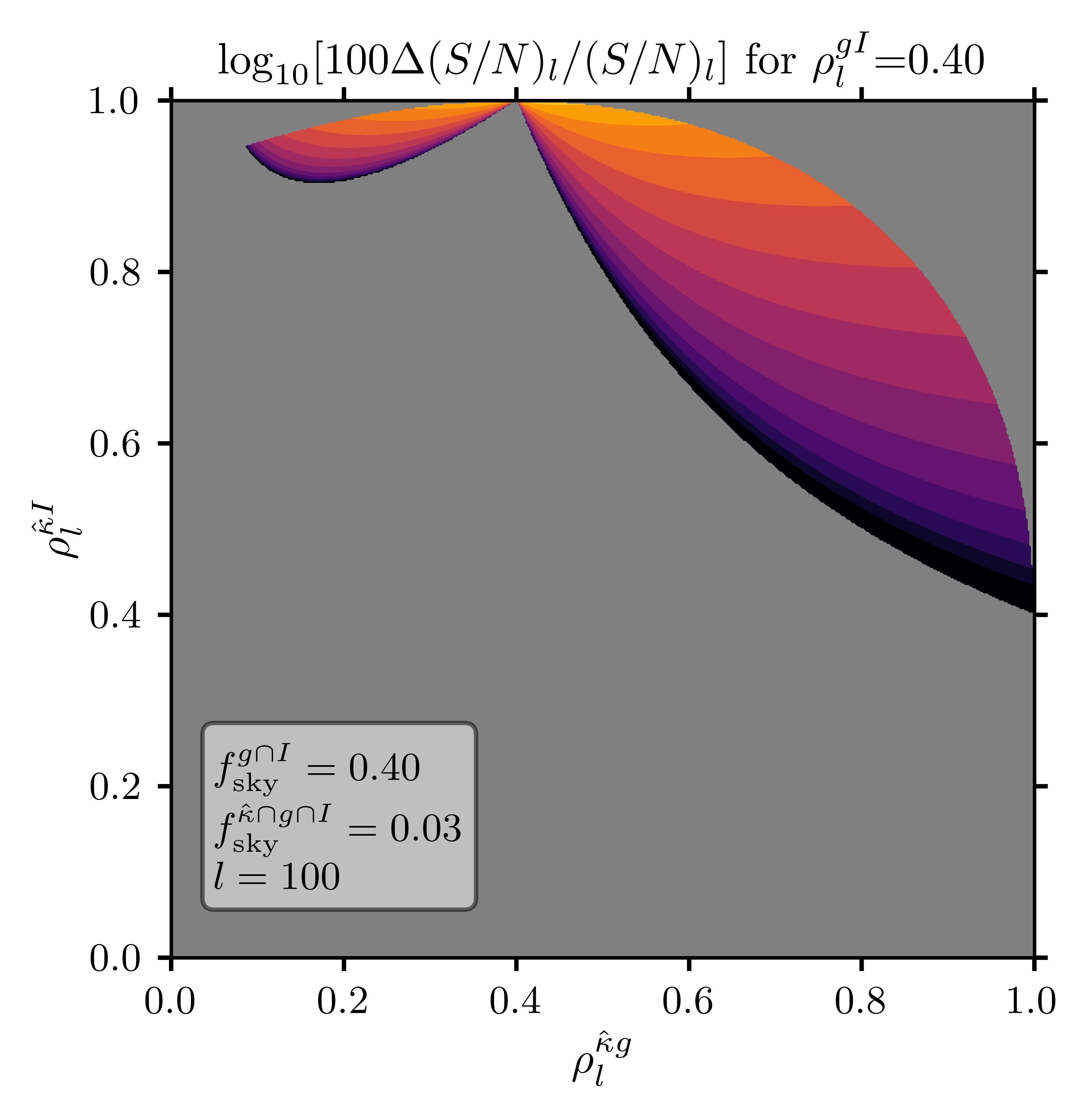}
    \includegraphics[width=0.32\textwidth]{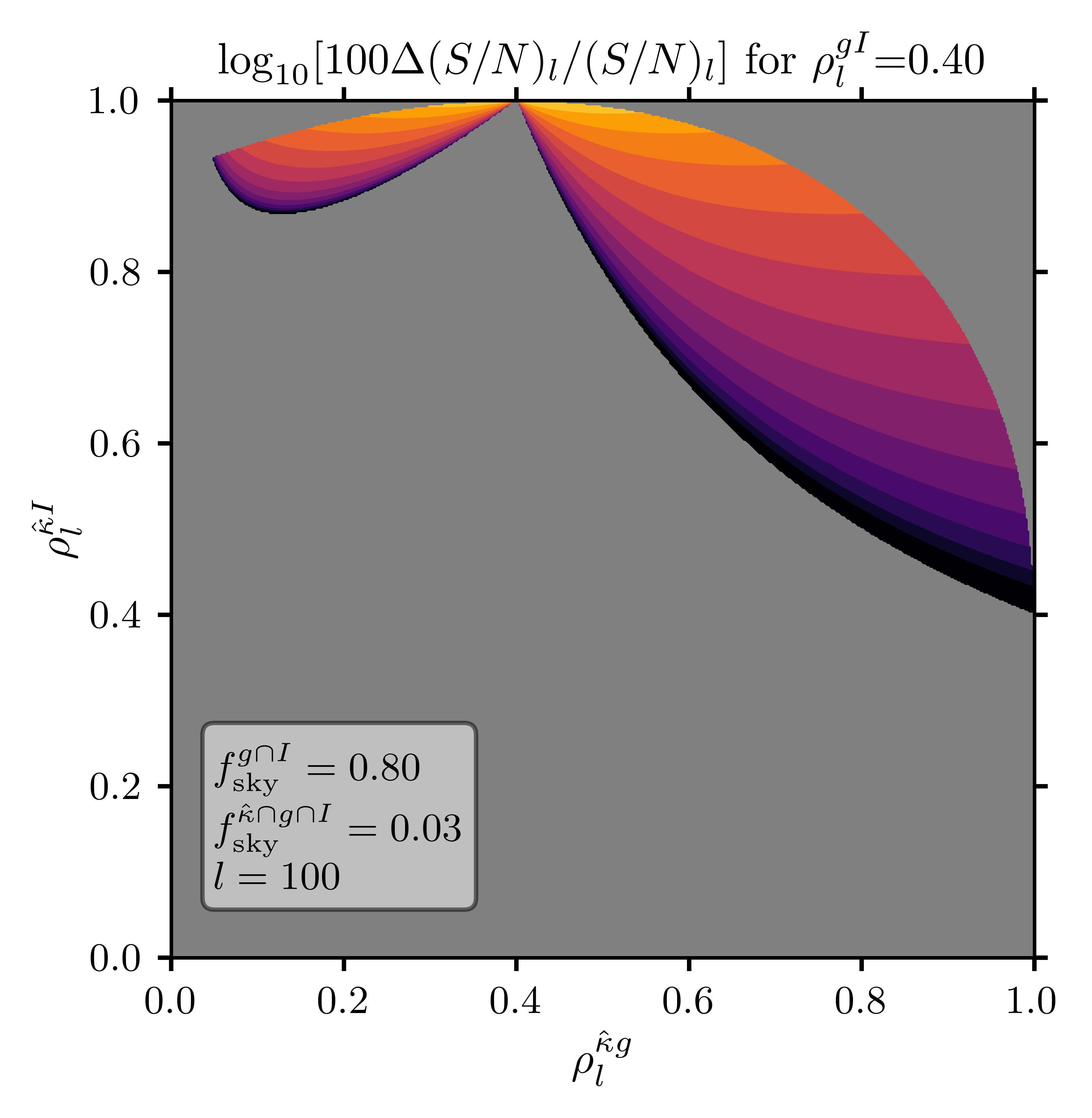}
    \caption{Impact of the size of the patch where $\hat{C}^{gI}_l$ is measured ($g \cap I$, disjoint from $\hat{\kappa} \cap g \cap I$) on the possible gains in S/N per mode of $C_l^{\hat{\kappa}g}$ when $\rhogI_l =0.4$, at the MSE optimum. (See text of appendix~\ref{appendix:generalization} for what scenarios these $f_{\mathrm{sky}}$ values  correspond to.) This is to be contrasted with figure~\ref{fig:StoN_gain_r=1}, which assumes a single patch with $f^{g \cap I}_{\mathrm{sky}}=f^{\hat{\kappa} \cap g \cap I}_{\mathrm{sky}}=0.03$ but the same color range given by the color bar in~\ref{fig:cbar_StoN}. Measuring $\hat{C}_l^{gI}$ on a patch of sky larger than that where we redshift-clean increases both the magnitude of gains in S/N and the domain where they are possible.}\label{fig:change_in_StoN_variable_r}
\end{figure}

Consider, on the other hand, the case where patches $\hat{\kappa} \cap g \cap I$ and $ g \cap I$ are disjoint. When this is the case, the covariance in the numerator of equation~\eqref{eqn:W_weights} can be ignored\footnote{Away from the lowest few multipoles.}. Furthermore, the precision with which  $\hat{C}^{gI}_l$ can be measured increases with the size of the $g \cap I$ patch; in the limit where the error in determining $C^{gI}_l$ is made small this way, $\mathcal{W}_l$ approaches unity as the signal can be accurately restored without adding significant amounts of variance. To state this more quantitatively, we once again look at the changes in S/N that can be achieved, this time specifying that the patches be disjoint\footnote{It is straight forward to verify that equations~\eqref{eqn:delta_S_to_N}, \eqref{eqn:delta_S_to_N_minimizing_MSE_joint} and~\eqref{eqn:delta_S_to_N_minimizing_MSE_disjoint} all equal each other when $\mathcal{W}_l=0$.}:
\begin{widetext}
\begin{align}\label{eqn:delta_S_to_N_minimizing_MSE_disjoint}
    \frac{\Delta (S/N)_l} {(S/N)_l} = & \left[1 - \left(1-\mathcal{W}_l\right) \gamma_l \rhokhatI_l \rhogI_l / \rhokhatg_l\right] \left[1 + \left(\rhokhatg_l\right)^{2}\right]^{\frac{1}{2}}\nonumber \\
    & \times \Bigg\{  1+\left(\rhokhatI_l\right)^2\gamma_l(\gamma_l-2) + \left(\rhokhatg_l-\gamma_l \rhogI_l \rhokhatI_l\right)^2 + \left(\mathcal{W}_l\gamma_l \rhokhatI_l\right)^2 \left(\frac{f_{\mathrm{sky}}^{\hat{\kappa} \cap g \cap I}}{f_{\mathrm{sky}}^{g \cap I}}\right) \left[1 + \left(\rhogI_l\right)^{2}\right] \Bigg\}^{-\frac{1}{2}}- 1\,.
\end{align}
\end{widetext}
Let us use this equation to assess the extent to which being able to measure $\hat{C}_l^{gI}$ on a larger, disjoint patch improves prospects for redshift cleaning. Suppose the CMB lensing reconstructions are obtained by a telescope on the South Pole while $g$ and $I$ are both measured on larger footprints by telescopes on the Atacama desert and/or space. For example, most of the 1500 $\mathrm{deg}^2$ ($f^{\hat{\kappa} \cap g \cap I}_{\mathrm{sky}}\approx 0.03$) covered by SPT-3G are contained within the 5000 $\mathrm{deg}^2$ observed by the Dark Energy Survey (DES)~\cite{ref:sevilla_noarbe_et_al_21} in its Y3 data release. If $g$ and $I$ are both drawn from DES observations, then $f^{g \cap I}_{\mathrm{sky}}\approx 0.09$  (after excising the region of overlap with $\hat{\kappa} \cap g \cap I$). More futuristically, if we assume that the 18000 $\mathrm{deg}^2$ coverage of LSST fully contains the SPT-3G patch, $f^{ g \cap I}_{\mathrm{sky}}\approx 0.4$. Finally, if $g$ and $I$ come from satellite observations on $80\%$ of the sky (we exclude regions near the Galactic plane), then $f^{ g \cap I}_{\mathrm{sky}}\approx 0.8$. Figure~\ref{fig:change_in_StoN_variable_r} shows the gain in S/N associated with these three scenarios for a fixed $\rhogI_l=0.4$ and $l=100$, so that they can be readily compared to figure~\ref{fig:StoN_gain_r=1}, for which a single patch with $f^{\hat{\kappa} \cap g \cap I}_{\mathrm{sky}} = f^{g \cap I}_{\mathrm{sky}}\approx 0.03$ is used. Notice that the prospects are significantly improved both in terms of the gain in S/N for fixed values of the correlation coefficients, and also in terms of the situations where gains are at all possible, as certain regions within the $\rhokhatI_l>\rhokhatg_l/\rho_l^{gI}$ branch now lend themselves to improved precision via redshift cleaning.

\section{Relating $\rho^{\kappa g}$,  $\rho^{\kappa I}$ and  $\rho^{ g I}$}\label{appendix:bound_on_rho}
In order to derive a relation between the correlation coefficients of tracers $\kappa$, $g$ and $I$, we will exploit an analogy between the cosine of the angle between vectors $\bm{X}$ and $\bm{Y}$,
\begin{equation}
    \cos \theta_{\bm{X}\bm{Y}} = \frac{\bm{X}\cdot \bm{Y}}{|\bm{X}||\bm{Y}|}\,,
\end{equation}
and the correlation coefficient between projected fields $X$ and $Y$ (we drop the multipole dependence for notational convenience),
\begin{equation}
    \rho^{XY} = \frac{C^{XY}}{\sqrt{C^{XX}} \sqrt{C^{YY}}}\,.
\end{equation}

Following this analogy, we promote our tracers to vectors in three-dimensional space, and seek an expression for $ \cos \theta_{\kappa g}$ in terms of $\cos \theta_{g I}$ and $\cos \theta_{\kappa I}$, and similarly for $\cos \theta_{\kappa I}$. Such a relation exists only if the three vectors are coplanar; fortunately, this limit suffices to place an upper bound on $ \cos \theta_{\kappa g}$, once we realize that 
for fixed $\theta_{g I}$ and $\theta_{\kappa I}$, the configuration that maximizes $\cos \theta_{\kappa g}$ is the one where all the vectors are coplanar. Consequently,
\begin{align}
    \rho^{\kappa g} & = \cos \theta_{\kappa g} \nonumber \\
    & \leq  \cos \left(\theta_{g I} - \theta_{\kappa I} \right) \nonumber\\
    & = \cos \theta_{g I} \cos \theta_{\kappa I} +  \sin \theta_{g I} \sin \theta_{\kappa I} \nonumber\\
    & = \rho^{g I }  \rho^{\kappa I } + \sqrt{\left[1-\left(\rho^{g I}\right)^2\right]\left[1-\left(\rho^{\kappa I}\right)^2\right]}\,,
\end{align}
where, in the second line, equality holds only when the vectors are coplanar. By the same logic,
\begin{equation}
    \rho^{\kappa I} \leq \rho^{g I }  \rho^{\kappa g } + \sqrt{\left[1-\left(\rho^{g I}\right)^2\right]\left[1-\left(\rho^{\kappa g}\right)^2\right]}\,.
\end{equation}

\section{Parametrizing tracers for forecasts}\label{appendix:parametrizing_tracers_forecasts}
In this section, we describe the tracer modelling underlying the forecasts of figure~\ref{fig:change_in_StoN}.

Our parametrization of the VRO LSST follows~\cite{ref:yu_17}. We assume the galaxies can be split into disjoint tomographic bins with edges at redshifts 0, 0.5, 1, 2, 3, 4 and 7. The first five bins ($0<z<4$) are then combined using the weights of~\cite{ref:sherwin_15} to more closely match the galaxy kernel to that of CMB lensing. This constitutes our tracer $I$ -- we also retain the last bin ($4<z<7$) to be used as our $g$ tracer. As per GNILC CIB and DESI BGS (see section~\ref{sec:data} for details about these tracers), our models come from theory-inspired fits to the data, once again following~\cite{ref:yu_17}.

On the CMB lensing side, we consider experiments with the characteristics of CMB-S4 and ACT~\cite{ref:darwish_21} DR6. We follow~\cite{ref:limitations_paper} in treating the CMB-S4 Gaussian lensing reconstruction noise as being white on signal-dominated scales. We take the white noise level to be $\Delta_{\kappa}=0.2$\,arcmin to match figure~3 of~\cite{ref:schmittfull_and_seljak}, which shows the reconstruction noise for a minimum-variance combination of temperature and polarization reconstructions -- the former up to $l^{T}_{\mathrm{max}}=3000$, the latter to $l^{E,B}_{\mathrm{max}}=3500$ and applied iteratively -- for a possible Stage-4 experiment with a symmetric, Gaussian beam with FWHM of 1 arcmin and 1 $\mu\mathrm{K\,arcmin}$ white noise. On the other hand, the noise curve for ACT DR6 was generated by assuming a 1.4 arcmin beam and a white noise level of $10\,\mu\mathrm{K\,arcmin}$.

\section{Modeling changes in spectra}\label{appendix:models_spectra_changes}
When the weights in equation~\eqref{eqn:mv_weights} are employed, the angular power spectrum of the redshift cleaned convergence map is
\begin{equation}
    C^{\hat{\kappa}^{\mathrm{cln}}\hat{\kappa}^{\mathrm{cln}}}_l = C^{\hat{\kappa} \hat{\kappa} }_l \left[1 - 2f_l\frac{C_l^{\kappa I}}{C_l^{\hat{\kappa}\hat{\kappa}}} + f_l^2\frac{ C_l^{I I}}{C_l^{\hat{\kappa}\hat{\kappa}}} \right]\,,
\end{equation}
such that the fractional change in lensing power after variance cancellation is
\begin{equation}\label{eqn:frac_auto_cls_general}
    \frac{\Delta C^{\hat{\kappa}\hat{\kappa}}_l}{C^{\hat{\kappa}\hat{\kappa}}_l} =  - 2f_l\frac{C_l^{\kappa I}}{C_l^{\hat{\kappa}\hat{\kappa}}} + f_l^2\frac{ C_l^{I I}}{C_l^{\hat{\kappa}\hat{\kappa}}}\,.
\end{equation}
If the fiducials are perfectly matched to the truth, these expressions simplify to
\begin{equation}\label{eqn:auto_cls}
    \hat{C}^{\hat{\kappa}^{\mathrm{cln}}\hat{\kappa}^{\mathrm{cln}}}_l = \hat{C}^{\hat{\kappa} \hat{\kappa} }_l \left[1 + \left(\rho_l^{\hat{\kappa} I}\right)^2 \gamma_l(\gamma_l-2)\right]\,,
\end{equation}
and
\begin{equation}
    \frac{\Delta C^{\hat{\kappa}\hat{\kappa}}_l}{C^{\hat{\kappa}\hat{\kappa}}_l} =  \left(\rhokhatI_l\right)^2 \gamma_l(\gamma_l-2)\,.
\end{equation}

On the other hand, the cross-correlation of $\hat{\kappa}^{\mathrm{cln}}$ with $g$ gives
\begin{equation}\label{eqn:cross_cls}
    C^{\hat{\kappa}^{\mathrm{cln}}g}_l = C^{\hat{\kappa} g}_l - f_{l} C_{l}^{g I}\,.
\end{equation}
If the fiducials match the truth, then
\begin{equation}
    C^{\hat{\kappa}^{\mathrm{cln}}g}_l = C^{\hat{\kappa} g}_l \left(1 - \gamma_l \rhogI_l \rhokhatI_l / \rhokhatg_l\right)\,.
\end{equation}
All in all, in the limit that the fiducials match the truth, the correlation of the cleaned convergence map with tracer $g$ becomes
\begin{equation}
    \rho^{\hat{\kappa}^{\mathrm{cln}} g}_l = \rho^{\hat{\kappa}g}_l\left[ \frac{1- \gamma_l \rhogI_l\rhokhatI_l/\rhokhatg_l}{\sqrt{1+(\rhokhatI_l)^{2} \gamma_l(\gamma_l-2)}}\right]\,.
\end{equation}
Furthermore, if $g$ and $I$ are completely uncorrelated, then
\begin{equation}
    \rho^{\hat{\kappa}^{\mathrm{cln}} g}_l = \frac{\rhokhatg_l}{\sqrt{1-\left(\rho_l^{\hat{\kappa} I}\right)^2}}\,.
\end{equation}

\section{Modeling variance reduction}\label{appendix:modeling_var}
We now model the change in variance after applying our method. In doing so, let us follow a route that highlights the different contributions to the variance of our estimator. We begin with the general expression for the variance of the difference of two correlated random variables,
\begin{align}
    \sigma^2 \left( \Clkhatdelg\right) - \sigma^2\left( \Clkhatg\right)  = & - \sigma^2 \left( \Clkhatdelg - \Clkhatg\right) \\
    & + 2 \sigma^2 \left( \Clkhatdelg\right) \\
    & - 2 \mathrm{Cov} \left(\Clkhatdelg,   \Clkhatg\right)\,.
\end{align}
Taking the fields to be Gaussian, we find that
\begin{align}\label{eqn:model_diff_of_var_over_var}
    \frac{\Delta \sigma^2 \left( \Clkhatg\right)}{\sigma^2\left( \Clkhatg\right)}   = & - \frac{\sigma^2 \left( \Delta \Clkhatg\right)}{\sigma^2\left( \Clkhatg\right)} \nonumber\\
    & + \frac{2 f_l \left(C_l^{\hat{\kappa}\hat{\kappa}}\right)^{-1}}{1 + \left(\rhokhatg_l\right)^{2}} \bigg[ \ClkhatI \left(f_l \frac{C_l^{II}}{C_l^{\hat{\kappa}I}}-1\right) \\
    & \hphantom{+ \frac{2 f_ls \left(C_l^{\hat{\kappa}\hat{\kappa}}\right)^{-1}}{1 + \left(\rhokhatg_l\right)^{2}} \bigg[a} f_l \frac{C_l^{gI}}{C_l^{gg}} \left(C_l^{gI} - C_l^{\hat{\kappa} g}\right)
    \bigg]\,.
\end{align}
%
Heuristically, the first term captures the variance reduction due to having removed structure common to $\hat{\kappa}$ and $I$. It can be modelled as
\begin{align}\label{eqn:model_var_of_diff_over_var}
    \frac{\sigma^2 \left( \Delta \Clkhatg\right)}{\sigma^2\left( \Clkhatg\right)} = f_l^{2} \frac{C_l^{II}}{C_l^{\hat{\kappa}\hat{\kappa}}} \left[\frac{1 + \left(\rhogI_l\right)^2}{1+\left(\rhokhatg_l\right)^2}\right]\,.
\end{align}
In the limit that $\gamma_l=1$ and the fiducial $C_{l}^{g I, \mathrm{fid}}$ and $C_{l}^{I I, \mathrm{fid}}$ used to build $f_l$ are well matched to the truth, the term in the second line of equation~\eqref{eqn:model_diff_of_var_over_var} is zero in the mean. The same goes for the term in the third line whenever $\rhogI=0$. In these circumstances -- which are very similar to some of the cases considered in the main text, where $\gamma_l=1$ by construction, and $\rhogI$ is small -- equation~\eqref{eqn:model_var_of_diff_over_var} becomes a less noisy estimator of the variance change.

\begin{figure}
    \centering
    \includegraphics[width=\columnwidth]{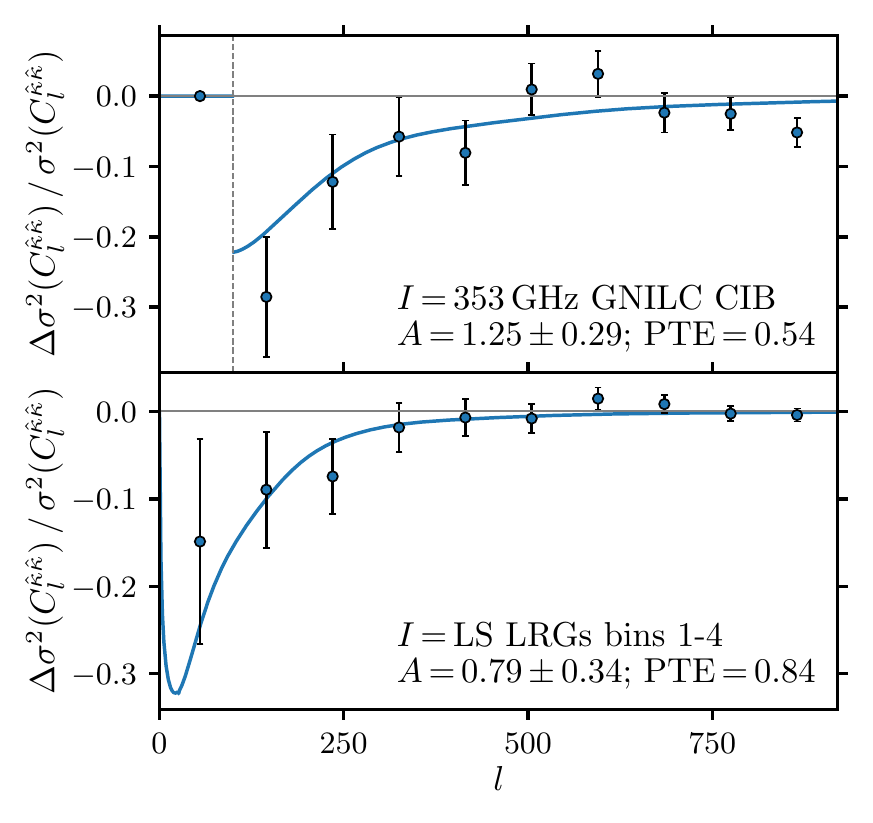}
    \caption{Reduction in variance of the CMB lensing bandpowers after cleaning a Planck PR4 $\hat{\kappa}$ map with a filtered tracer $I$ (setting $\gamma_l=1$). Solid lines show the fiducial models with $A=1$, while the best-fit values are annotated in each panel.}
    \label{fig:clkk_var_change}
\end{figure}

Let us also quote, for completeness, the expected fractional change in variance of the convergence bandpowers:
\begin{align}\label{eqn:clkk_variance}
    \frac{\sigma^2 \left( \Delta C_l^{\hat{\kappa}\hat{\kappa}} \right)}{\sigma^2\left( C_l^{\hat{\kappa}\hat{\kappa}} \right)} =  \left( 1 - 2f_l\frac{\ClkhatI}{C_l^{\hat{\kappa}\hat{\kappa}}} + f_l^2 \frac{C_l^{II}}{C_l^{\hat{\kappa}\hat{\kappa}} } \right)^2 - 1\,.
\end{align}
If the fiducial spectra are accurate, this gives
\begin{align}\label{eqn:clkk_variance_simplified}
    \frac{\sigma^2 \left( \Delta C_l^{\hat{\kappa}\hat{\kappa}} \right)}{\sigma^2\left( C_l^{\hat{\kappa}\hat{\kappa}} \right)} = \left(\rhokhatI_l\right)^2 \gamma_l \big[& -4 +2\gamma_l \nonumber \\
    & + \left(\rhokhatI_l\right)^2 \gamma_l \left[4(1-\gamma_l) + \gamma_l\right]\big] \,.
\end{align}
In figure~\ref{fig:clkk_var_change}, we show with actual data that the variance of estimates of the convergence bandpowers is reduced after redshift cleaning. The data we work with are the same that went into producing figure~\ref{fig:delta_ckk}: we clean Planck PR4 lensing with the GNILC CIB map, or with the combination of all four LRG bins. To gauge the statistical significance of this demonstration, we proceed as in section~\ref{sec:results} and fit the data with a rescaled version of the model in equation~\eqref{eqn:clkk_variance_simplified}. The best-fit values for this amplitude parameter $A$ are quoted in the figure.  Thus analyzed, variance cancellation is detected with approximately $4.3\sigma$ confidence when cleaning with the CIB, and $2.3\sigma$ with LS LRGs. The models provide good fits to the data, as evidenced by the PTE values.
\bibliography{main}

\end{document}